\documentclass{dnce}
\usepackage{dnce}
\usepackage{fixltx2e}
\Page{1}

\usepackage{units}
\usepackage{epstopdf}
\theoremstyle{definition}
\theoremstyle{remark}
\theoremstyle{plain}

\theoremstyle{remark}

\theoremstyle{definition}

\titlemark{E.F.D. Evangelista, \emph{et al.}/ Discontinuity, Nonlinearity, and Complexity Vol-number (year) page1--page2}
\authormark{E.F.D. Evangelista, \emph{et al.}/ Discontinuity, Nonlinearity, and Complexity Vol-number (year) page1--page2}

\begin{document}

\title{\first{Simulating the Interaction of a Comet With the Solar Wind Using a \newline Magnetohydrodynamic Model\newline \small{\textbf{This paper was published at}: \textit{Discontinuity, Nonlinearity, and Complexity} \textbf{7}(2), 143-149 (2018)}.\newline \small{DOI:10.5890/DNC.2018.06.003}}}

\setcounter{footnote}{1}

\author{\noindent\large Edgard de F. D. Evangelista$^1$\footnote{Corresponding author.\\Email address:
edgard.freitas.diniz@gmail.com}, Margarete O. Domingues$^2$, Odim Mendes$^3$, Oswaldo D. Miranda$^4$}

\address{$^1$\normalsize Laborat\'{o}rio Associado de Computa\c{c}\~{a}o e Matem\'{a}tica Aplicada, PCI/MCTI/INPE, S\~{a}o Jos\'{e} dos Campos, Brazil \\
$^2$\normalsize Laborat\'{o}rio Associado de Computa\c{c}\~{a}o e Matem\'{a}tica Aplicada, INPE, S\~{a}o Jos\'{e} dos Campos, Brazil \\
$^3$\normalsize Divis\~{a}o de Geof\'{i}sica Espacial, INPE, S\~{a}o Jos\'{e} dos Campos, Brazil\\
$^4$\normalsize Divis\~{a}o de Astrof\'{i}sica, INPE, S\~{a}o Jos\'{e} dos Campos, Brazil }

\abstract{
\begin{table}[h!]
\vspace*{-5mm}
\doublerulesep 0.05pt
\tabcolsep 7.8mm
\vspace*{2mm}
\setlength{\tabcolsep}{7.5pt}
\hspace*{-2.5mm}\begin{tabular*}{16.5cm}{r|||||l}
\multicolumn{2}{l}{\rule[-6pt]{16.5cm}{.01pt}}\\
\parbox[t]{6cm}{\small
\vspace*{.5mm}
\hfill {\bf Submission Info}\par
\vspace*{2mm}
\hfill Communicated by Referees\par
\hfill Received DAY MON YEAR \par
\hfill Accepted DAY MON YEAR\par
\hfill Available online DAY MON YEAR\par
\noindent\rule[-2pt]{6.3cm}{.1pt}\par
\vspace*{2mm}
\hfill {\bf Keywords}\par
\vspace*{2mm}
\hfill Magnetohydrodynamics\par
\hfill Computational Model\par
\hfill Instabilities\par
\hfill Comet}
&
\parbox[t]{9.85cm}{
\vspace*{.5mm}
{\normalsize\bf Abstract}\par
\renewcommand{\baselinestretch}{.8}
\normalsize \vspace*{2mm} {\small We present simulations of a comet interacting with the solar wind. Such simulations are treated in the framework of the ideal, 2D magnetohydrodynamics (MHD), using the FLASH code in order to solve the equations of such a formalism. Besides, the comet is treated as a spherically symmetric source of ions in the equations of MHD. We generate results considering several scenarios, using different values for the physical parameters of the solar wind and of the comet in each case. Our aim is to study the influence of the solar wind on the characteristics of the comet and, given the nonlinear nature of the MHD, we search for the occurrence of phenomena which are typical of nonlinear systems such as instabilities and turbulence.}
\par
\par
\par
\hfill{\scriptsize\copyright 2012 L\&H Scientific Publishing, LLC. All rights reserved.}}\\[-4mm]
&\\
\multicolumn{2}{l}{\rule[15pt]{16.5cm}{.01pt}}\\\end{tabular*}
\vspace*{-7mm}
\end{table}
}

\maketitle
\thispagestyle{first}

\renewcommand{\baselinestretch}{1}
\normalsize

\section{Introduction}

\noindent
A comet may be defined as a object formed by rock, dust and water ice and which undergoes the action of the gravitational field of Sun. In such objects it is usual the presence of gases such as ammonia, methane and carbon oxides\cite{greenberg:1998}.

Concerning the physical models, it is common to figure a comet as a ball of ``dirty ice''\cite{carroll:2007}, which undergoes sublimation when it reachs the regions close to the perihelion of its orbit. The sublimation causes the ejection of gases and dust, forming the characteristic coma and tail of the comet.

The simulation of the interaction between the coma and tail with the solar wind are important once it provides physical insights on the processes characteristic of cometary atmospheres. Such simulations may yield information about the physical and chemical processes in the coma. Besides, the interaction of the comet with the plasma of the solar wind may be used as a tool for tracing the behaviour of the space environment.

Given the nonlinear nature of the MHD, we should expect that scenarios such that the ones simulated in this paper will present characteristics wich are typical of nonlinear systems such as instabilities and turbulence. Therefore, the simulations shown here are useful for the study of the formation of such phenomena in cometary atmospheres.

This work consists in the numerical simulation of a comet interacting with the solar wind, using the formalism of the ideal, two-dimensional MHD. In such simulations the comet is modelled by inserting a spherically isotropic source of ions in the system of equations of MHD. We use the FLASH Code of the University of Chicago\cite{fryxell:2000} in order to generate the results.

Our aim here is to study the influence of the physical characteristics of the solar wind on the behaviour of the coma and tail and search for MHD instabilities in this scenario. In order to do so, we perform simulations using different values for the magnetic field and velocity of the solar wind, besides considering different values for the gas production rate $Q$ of the comet. More specifically, we consider the values of $\unit[1.0]{nT}$, $\unit[3.5]{nT}$ and $\unit[6.0]{nT}$ for the component $B_z$ of the magnetic field; the values of $\unit[4.0\times 10^{-7}]{cm~s^{-1}}$, $\unit[5.0\times 10^{-7}]{cm~s^{-1}}$ and $\unit[6.0\times 10^{-7}]{cm~s^{-1}}$ for the velocity of the solar wind; and $\unit[1.0\times 10^{28}]{s^{-1}}$, $\unit[1.0\times 10^{29}]{s^{-1}}$ and $\unit[1.0\times 10^{30}]{s^{-1}}$ for the values of $Q$. In each simulation we use particular a combination of $B_z$, $v$ and $Q$, covering all the possible combinations.

This paper is organized in the following way: in Section 2 we present the basic formalism of MHD, showing how the terms representing the comet are included in the equations; in Section 3 we present the numerical aspects of the simulations, such as the computational parameters; in Section 4 we present the results and discussions. Further, the conclusions are shown in Section 5.

\section{The MHD model}

\noindent
The MHD formalism describes the interaction between compressible plasma and magnetic fields. This model is built when one combines the equations governing the fluid dynamics with Maxwell's equations of the electromagnetism resulting, in a 3D domain, in a systems of eight equations describing the conservation of mass and momentum, Faraday's law and the conservation of energy\cite{powell:1999}.
In the present work we are considering the ideal MHD, that is, the case where dissipative terms such as resistivity and viscosity are neglected. Besides, we restrict ourselves to the non-relativistic case, that is, all the velocitites considered are small when compared to the speed of light in vacuum.
The system of equations of the MHD is given by\cite{ekenback:2008}:

\begin{equation}
\label{mhd1}
    \frac{\partial}{\partial t}\left( \begin{array}{c}
    \rho \\
    \rho \boldsymbol{v} \\
    \boldsymbol{B}\\
    \rho E
    \end{array} \right)+\nabla\cdot
    \mathbb{F}=
    \left( \begin{array}{c}
    A \\
    -\boldsymbol{v}A \\
    \boldsymbol{0} \\
    Au_{n}^{2}/2
     \end{array} \right),
  \end{equation}
where
\begin{equation}
\label{mhd2}
    \mathbb{F}=\left( \begin{array}{c}
    \rho\boldsymbol{v} \\
    \rho\boldsymbol{v}\boldsymbol{v}-\boldsymbol{B}\boldsymbol{B}+\left(p+B^{2}/2\right)\mathbb{I} \\
    \boldsymbol{v}\boldsymbol{B}-\boldsymbol{B}\boldsymbol{v} \\
    \boldsymbol{v}\left(\rho E+p+B^{2}/2\right)-\boldsymbol{B}(\boldsymbol{v}\boldsymbol{B})
    \end{array} \right).
\end{equation}

In Eq.~(\ref{mhd1}) and (\ref{mhd2}), $\rho$ is the mass density, $p$ is the thermal pressure, $\boldsymbol{v}$ is the velocity, $\boldsymbol{B}$ is the magnetic field, $\mathbb{I}$ is  the 3x3 identity matrix, $u_{n}$ is the radial velocity in which the gas is ejected from the comet and $E$ is the total energy. On the other hand, $A$ is the source term representing the comet, which has the form\cite{ekenback:2008,gunell:2015}:

\begin{equation}
  \unit[A=A(r)=\frac{Q}{4\pi\lambda r^{2}}e^{-r/\lambda}]{m^{-3}~s^{-1}},
\label{sourceterm}
\end{equation}
in which $r$ is the radial distance from the center of the comet, $Q$ is the gas production rate (given in particles per second) and $\lambda=\tau u_{n}$ is the typical ionisation distance, where $\tau=\unit[3.03\times 10^{5}]{s}$ is the typical ionisation time. Besides, we are considering the gas is ejected from the comet at a radial velocity $u_{n}$ of $\unit[1.0\times 10^{5}]{cm~s^{-1}}$\cite{young:2004}.

\section{Numerical methods}

\noindent
The simulations are performed with the FLASH Code distributed by the Center for Astrophysical Thermonuclear Flashes at the University of Chicago\cite{fryxell:2000,dubey:2008}. This code permits the user to choose several parameters of the problem being simulated, such as initial and boundary conditions and the schemes for the calculation of the numerical fluxes, for example.

In the present simulations it is used the Roe flux and a second order Godunov method called piecewise-parabolic (PPM) algorithm. Besides, it is used the adaptive refinement scheme known as PARAMESH.

The 2D domain used here have the size $x\in [-2.8, 1.6] \unit[\times 10^{11}]{cm}$ and 
$y\in [-2.2, 2.2] \unit[\times 10^{11}]{cm}$. The boundary conditions are: \emph{outflow} at x\textsubscript{\scriptsize{left}}; \emph{user defined} at x\textsubscript{\scriptsize{right}}; and \emph{periodic} at y\textsubscript{\scriptsize{left}}  and  y\textsubscript{\scriptsize{right}}.
The \emph{outflow} boundary condition stands for a null gradient, that is, the waves can leave the domain, while the \emph{periodic} condition can be figured as a ``wrap-around'' of the domain. The \emph{user defined} condition is the one defined by the user and here represents the solar wind emerging from x\textsubscript{\scriptsize{right}} and flowing in the negative x-direction. We use a CFL\cite{courant:1967} condition of $0.8$ and adiabatic index $\gamma=5/3$. The final time of the simulation is $\unit[1.0\times 10^{5}]{s}$.

The simulation starts with a block of $8\times8$ cells and evolves up to six levels of refinement. Each level of refinement quadruplicates the number of blocks such that at the end of the process the domain would have $32\times32$ blocks if it would be entirely refined. This case correspond to a mesh of $256\times256$ cells.

Table~\ref{Tab1} shows the numerical values of the magnetic field $\boldsymbol{B}$, the velocity of the solar wind $\boldsymbol{v}$ and the gas production rate $Q$ to be inserted in the simulations. Here, $(B_x, B_y, B_z)$ are the components of the magnetic field at the position of the comet and $(v_x,v_y,v_z)$ are the components of the velocity of the solar wind. We are considering the GSE (Geocentric Solar Ecliptic) coordinate system.

The values for $\boldsymbol{B}$ and $\boldsymbol{v}$ are consistent with the ones found in \cite{lebedev:2015} and with the data extracted from the GSFC/SPDF OMNIWeb interface\footnote{\url{http://omniweb.gsfc.nasa.gov/form/omni_min.html}}, which provides data collected by the ACE, Wind, IMP 8 and Geotail spacecrafts. The maximum value for $Q$ is the one used in \cite{ekenback:2008}, while the minimum value is near to the maximum one found in \cite{lebedev:2015}. Further, the temperature has the fixed value of $T=\unit[2\times 10^{5}]{K}$.

\begin{table}[h!]
\doublerulesep 0.1pt
\tabcolsep 7.8mm
\centering
\caption{\rm Parameters of the comet and of the solar wind. The rate $Q$ is given in particles per second.}
\vspace*{2mm}
\renewcommand{\arraystretch}{1.3}
\setlength{\tabcolsep}{20pt}
\footnotesize{\begin{tabular*}{9.6cm}{cccc}
\hline\hline\hline
 & $Q$ & $(B_x, B_y, B_z)$ & $(v_x,v_y,v_z)$ \\
 &
$\unit[10^{28}]{s^{-1}}$ &
$\unit[]{nT}$ &
$\unit[10^{7}]{cm~s^{-1}}$ \\ \hline
 $1$ & $1.0$ & $(0,0,1.0)$ & $(-4.0,0,0)$ \\
 $2$ & $10.0$ & $(0,0,3.5)$ & $(-5.0,0,0)$ \\
 $3$ & $100.0$ & $(0,0,6.0)$ & $(-6.0,0,0)$ 
\\\hline\hline\hline\end{tabular*}
}
\renewcommand{\arraystretch}{1}
\label{Tab1}
\end{table}

\section{Simulations}

\begin{figure}
\centering
\begin{tabular}{ccc}
 \includegraphics[width=0.305\linewidth]{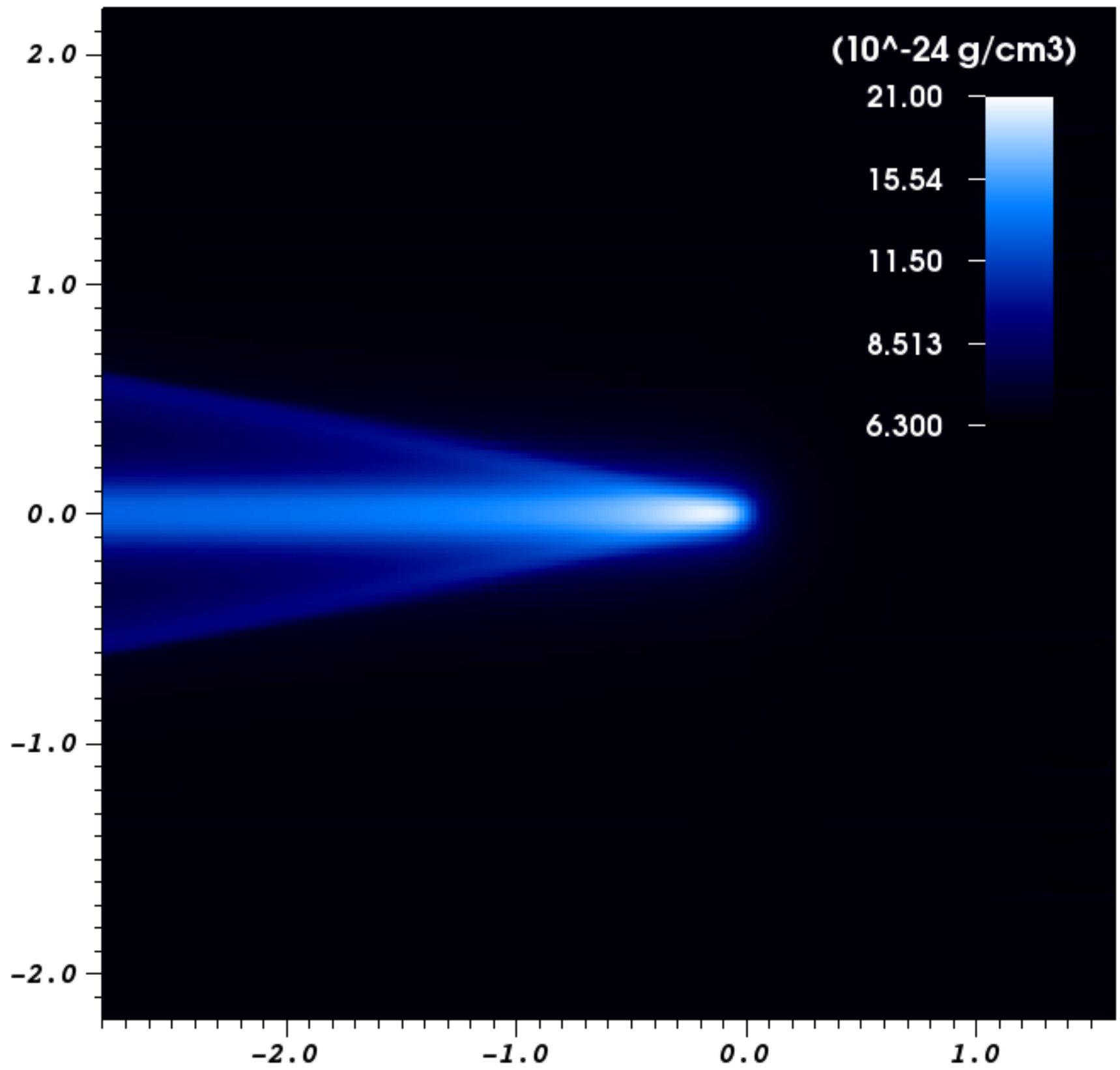} & 
 \includegraphics[width=0.305\linewidth]{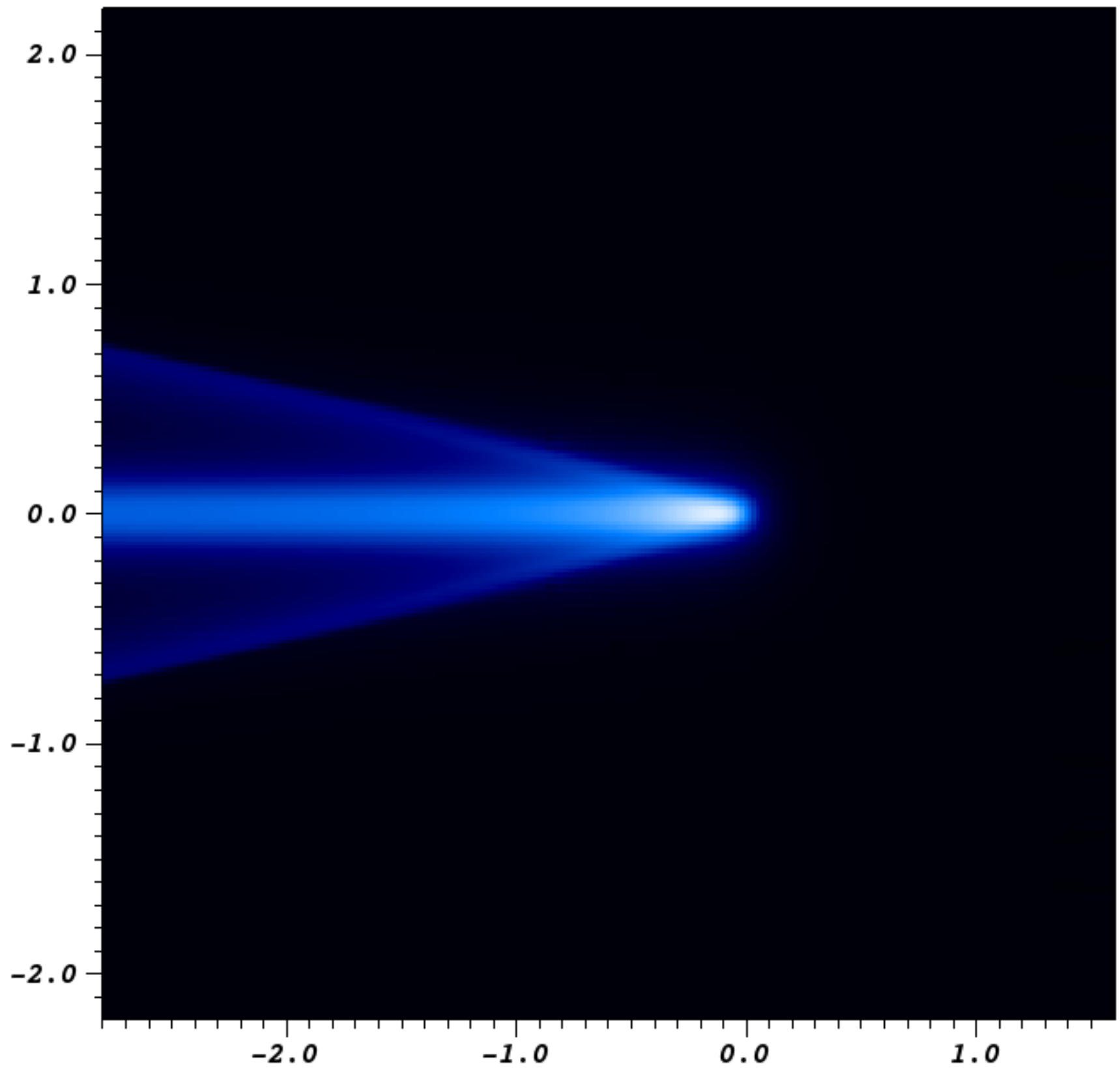} &
 \includegraphics[width=0.305\linewidth]{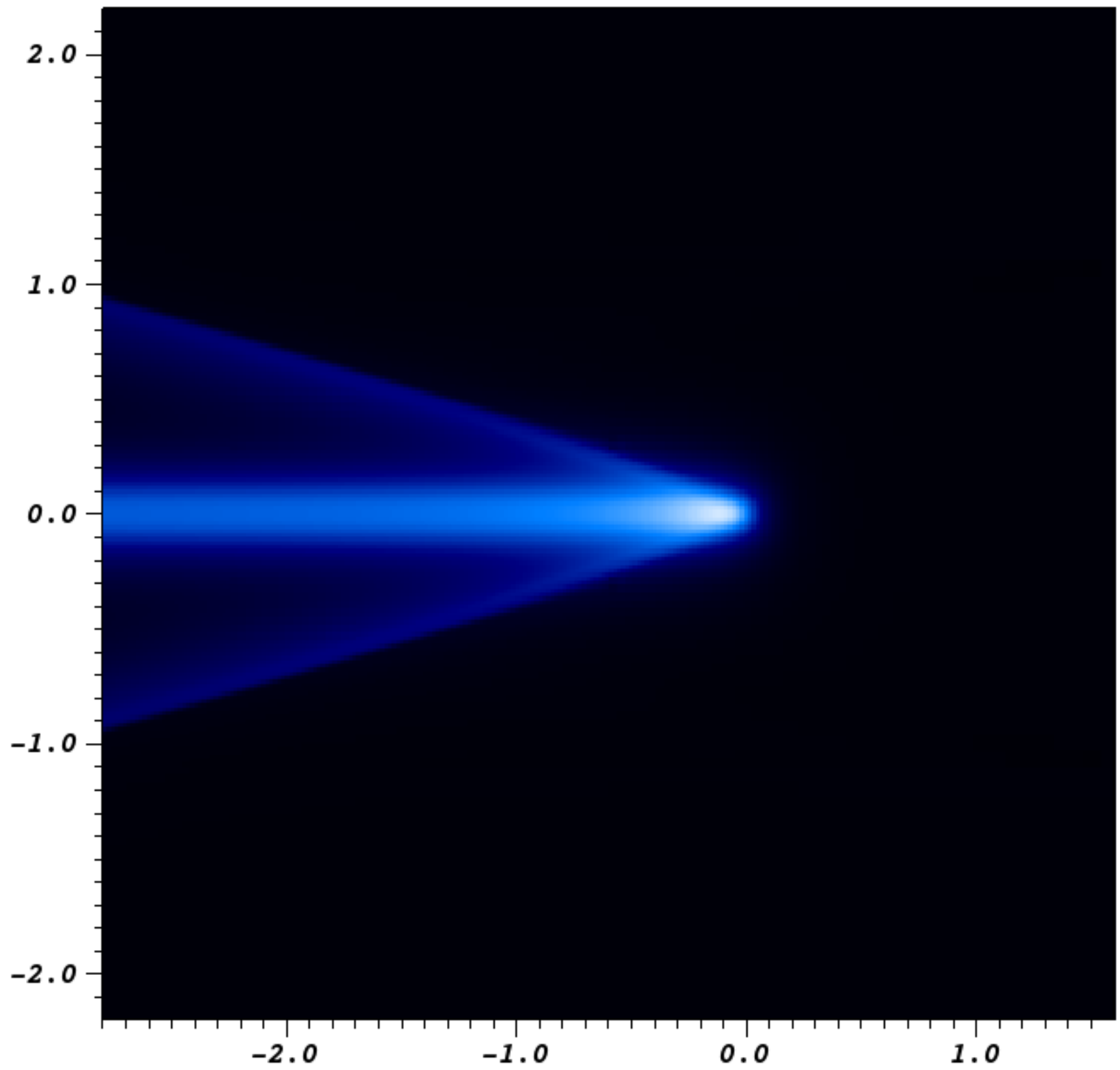} \\
 
 \includegraphics[width=0.305\linewidth]{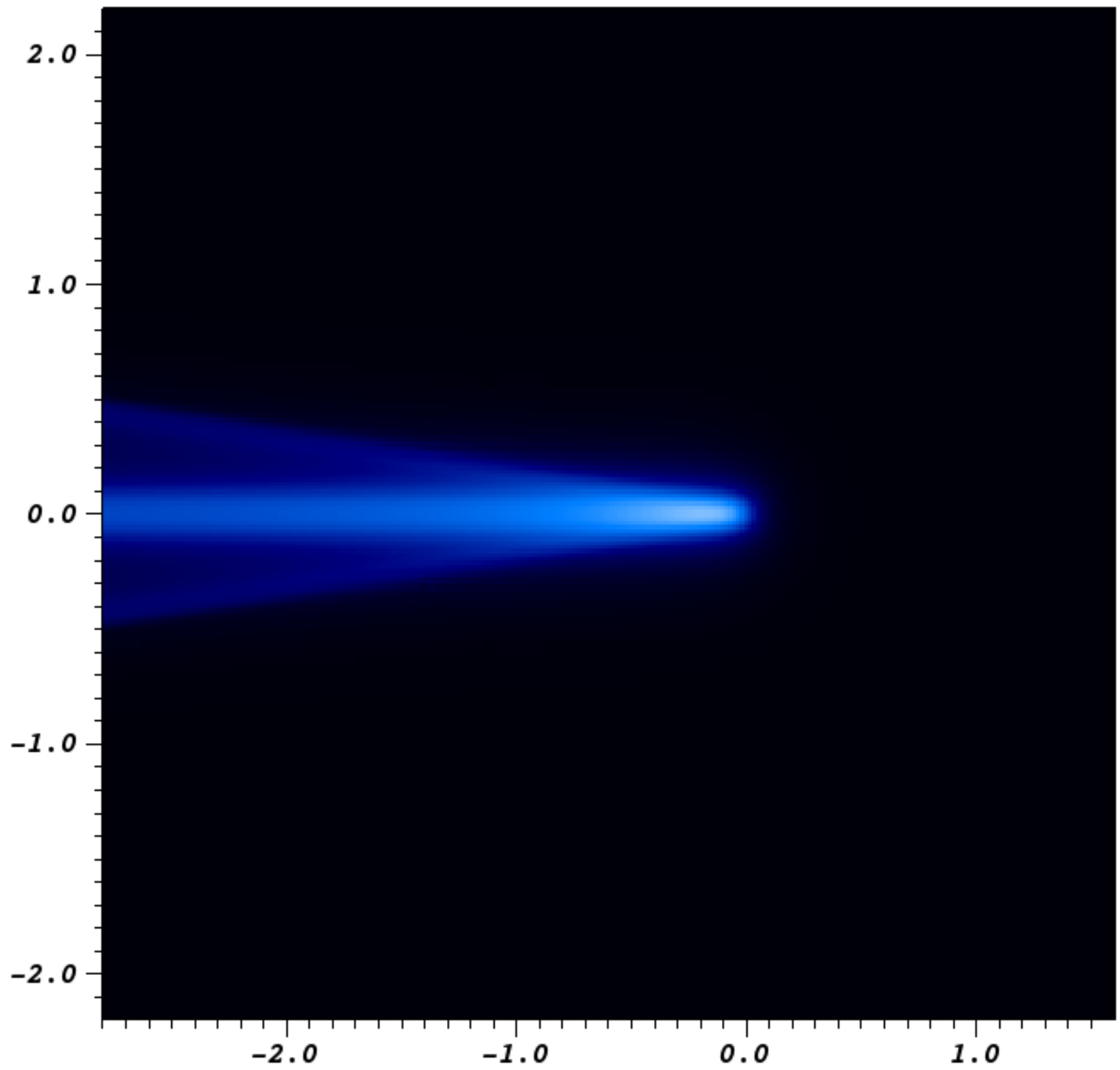} & 
 \includegraphics[width=0.305\linewidth]{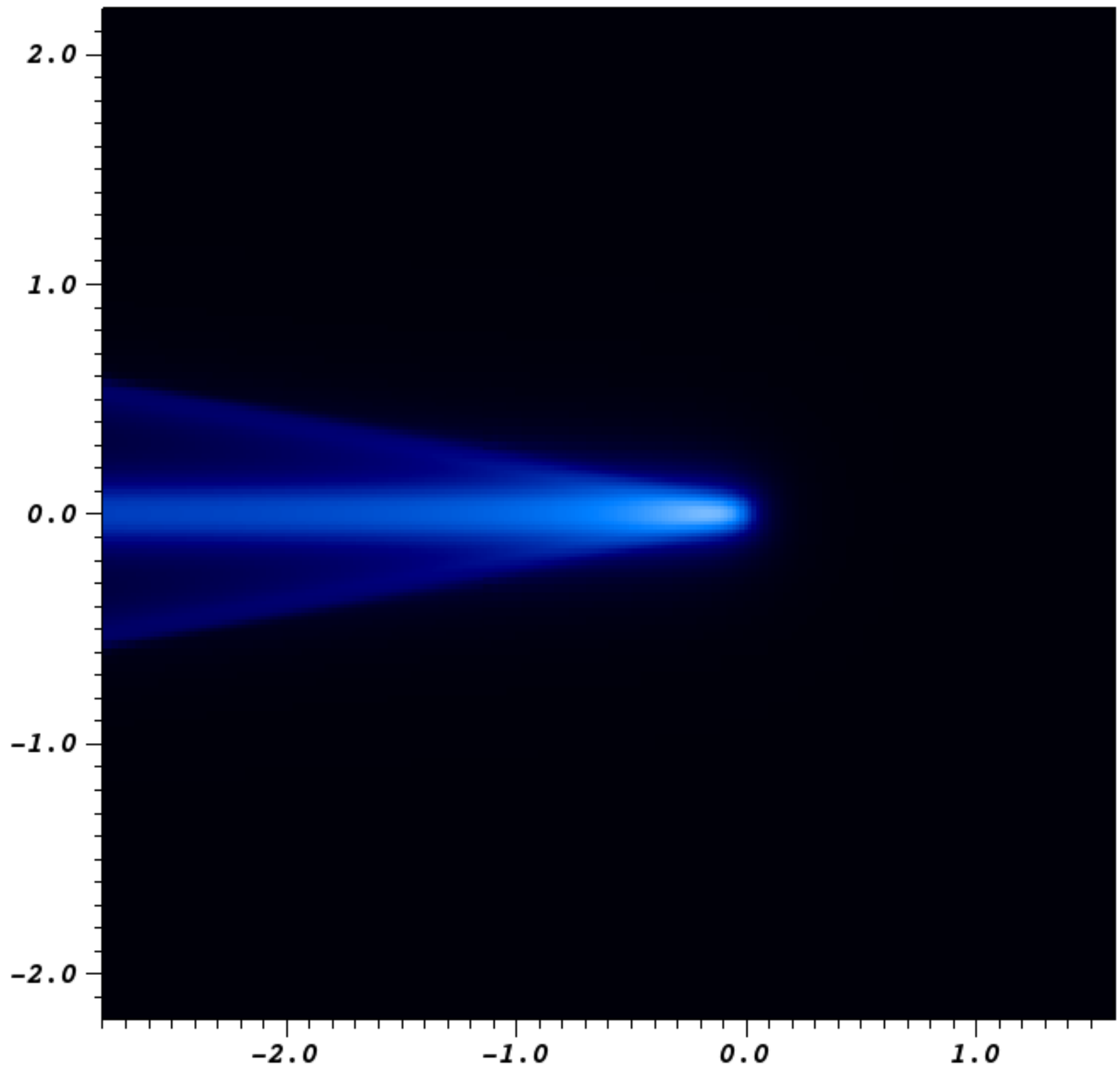} &
 \includegraphics[width=0.305\linewidth]{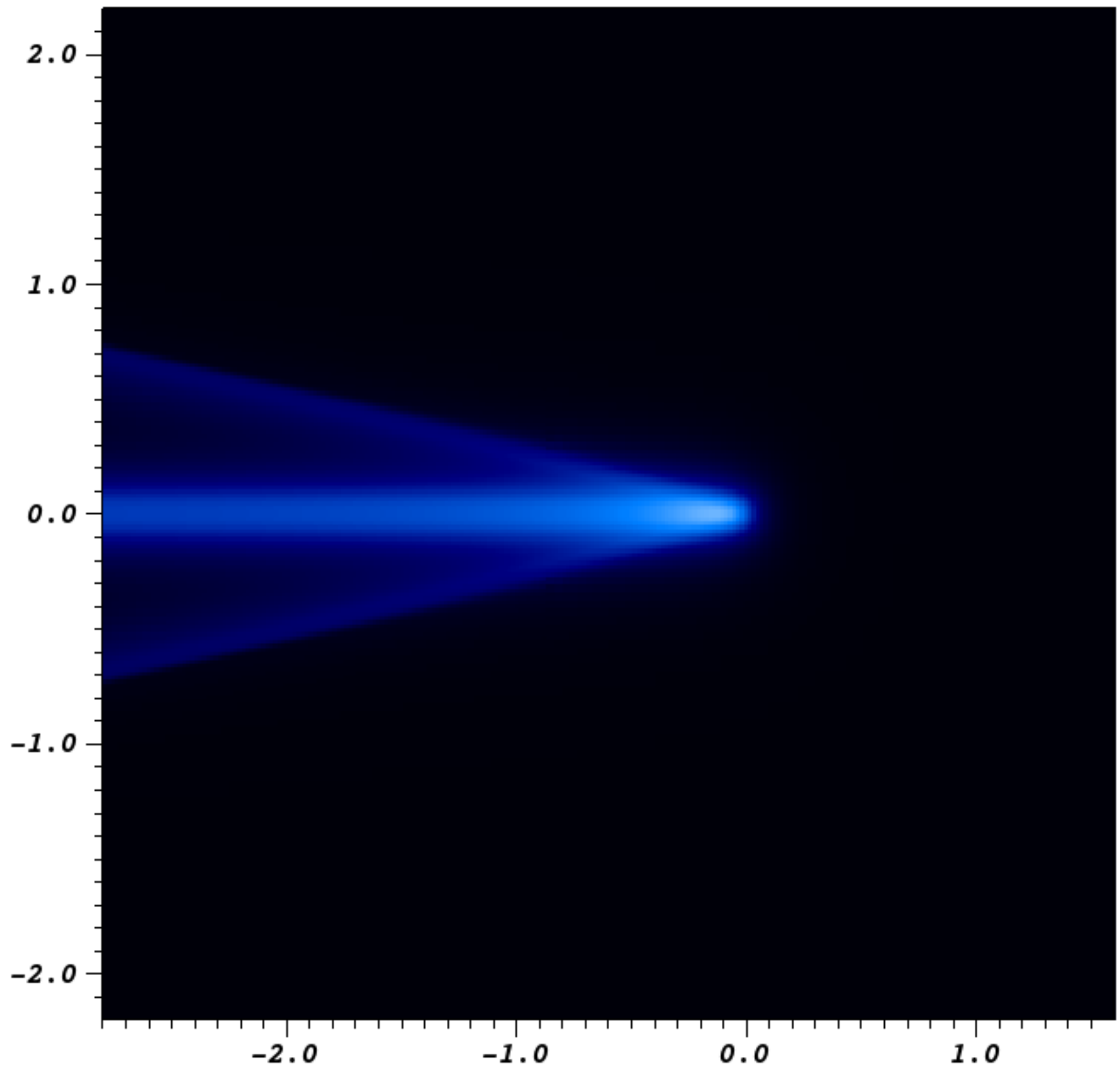} \\
 
 \includegraphics[width=0.305\linewidth]{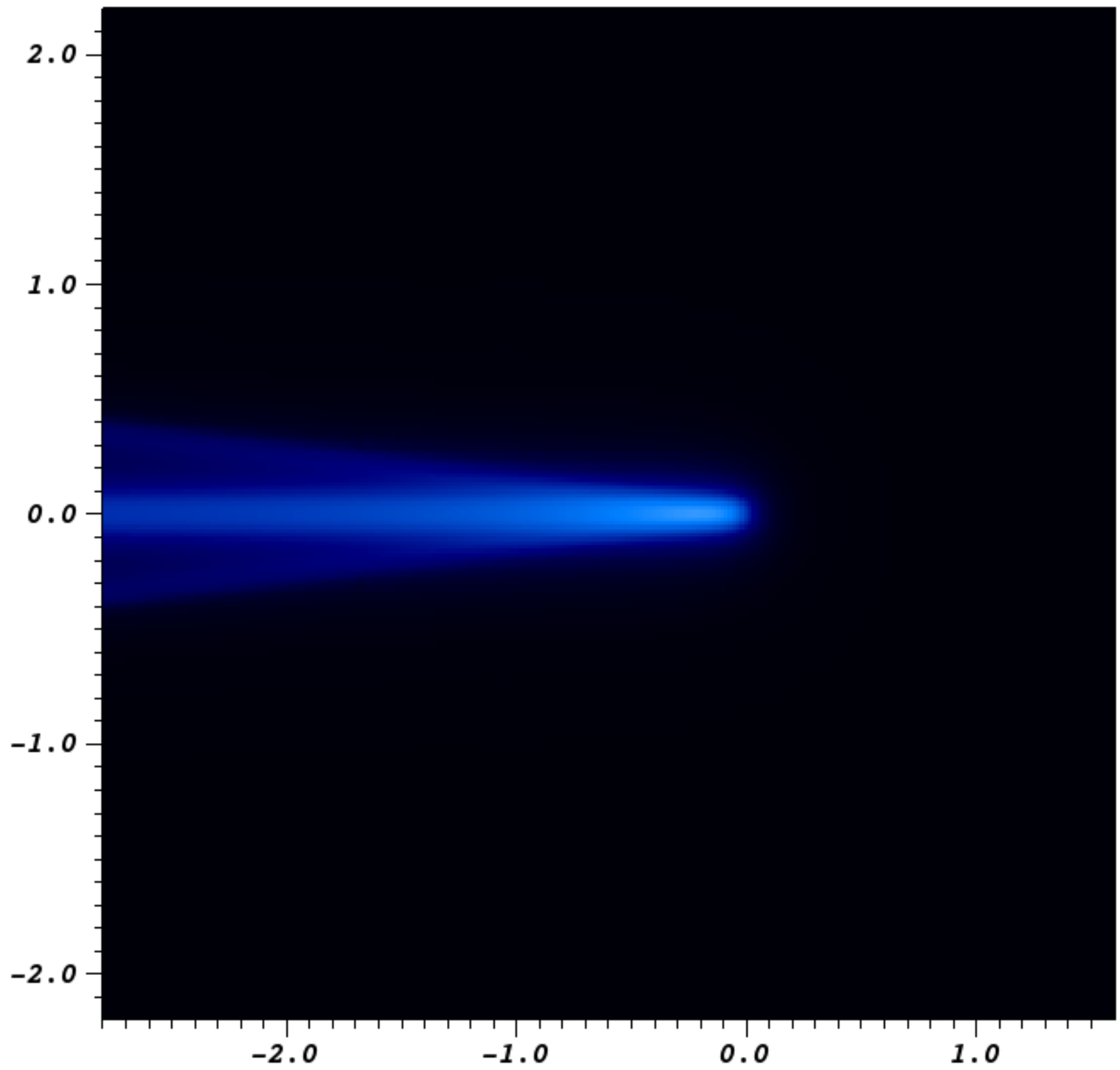} &
 \includegraphics[width=0.305\linewidth]{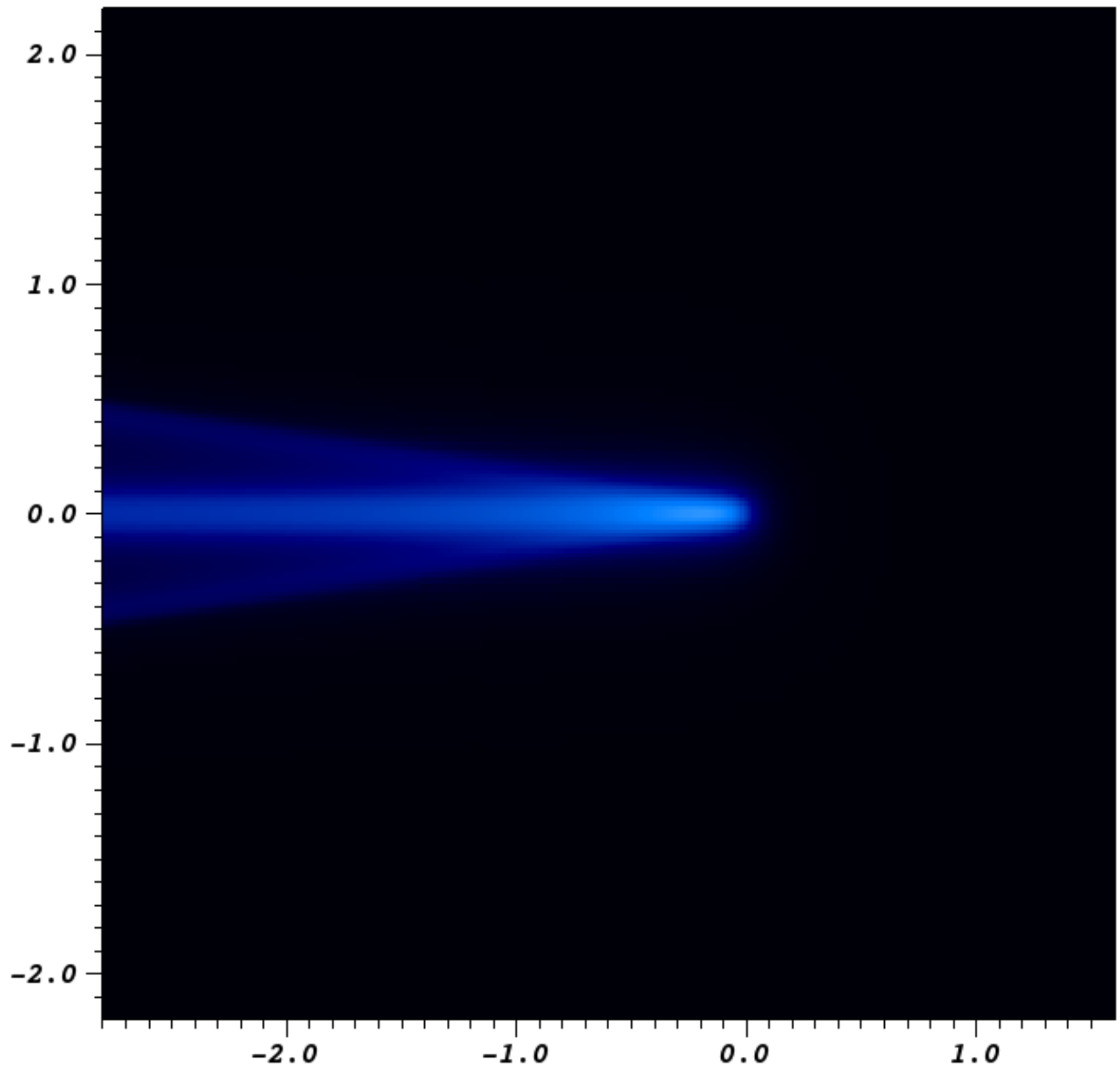} &
 \includegraphics[width=0.305\linewidth]{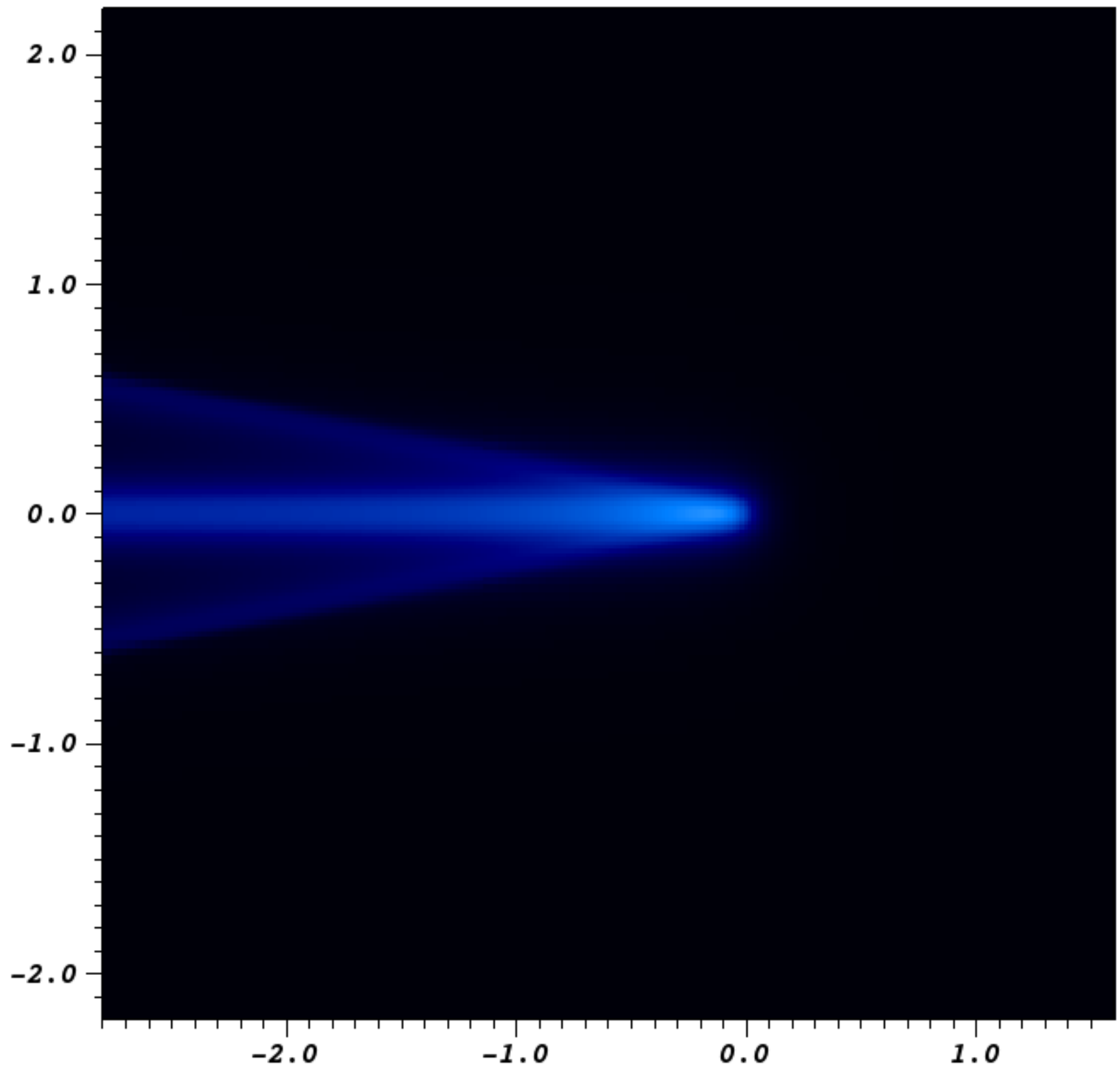} \\
\end{tabular}
\caption{Panel with the density profiles for $\unit[Q=1.0\times 10^{28}]{s^{-1}}$. The first, second and third lines of the panel were generated with $\boldsymbol{v}$ given in the lines 1,2 and 3 of Table~\ref{Tab1}, respectively. The first, second and third columns were generated using $\boldsymbol{B}$ given in the lines 1,2 and 3 of Table~\ref{Tab1}, respectively. The upper left profile shows the color bar with the values of the density. The length scale is $\unit[10^{11}]{cm}$.}
\label{figsQ1}
\end{figure}

\noindent
We performed the simulations using all the possible combinations of the parameters given in Table~\ref{Tab1}. Figure~\ref{figsQ1} shows the panel formed by the density profiles of the simulations for $\unit[Q=1.0\times 10^{28}]{s^{-1}}$. The first, second and third lines of such a panel correspond to the velocities of the solar wind given in the lines 1, 2 and 3 of Table~\ref{Tab1}, respectively. On the other hand, the first, second and third columns of Fig.~\ref{figsQ1} corresponds to the magnetic fields given in the lines 1, 2 and 3 of Table~\ref{Tab1}, respectively. The upper left profiles in Figs.~1-3 show the color bars with the scale of values for the density.

\begin{figure}
\centering
\begin{tabular}{ccc}
 \includegraphics[width=0.305\linewidth]{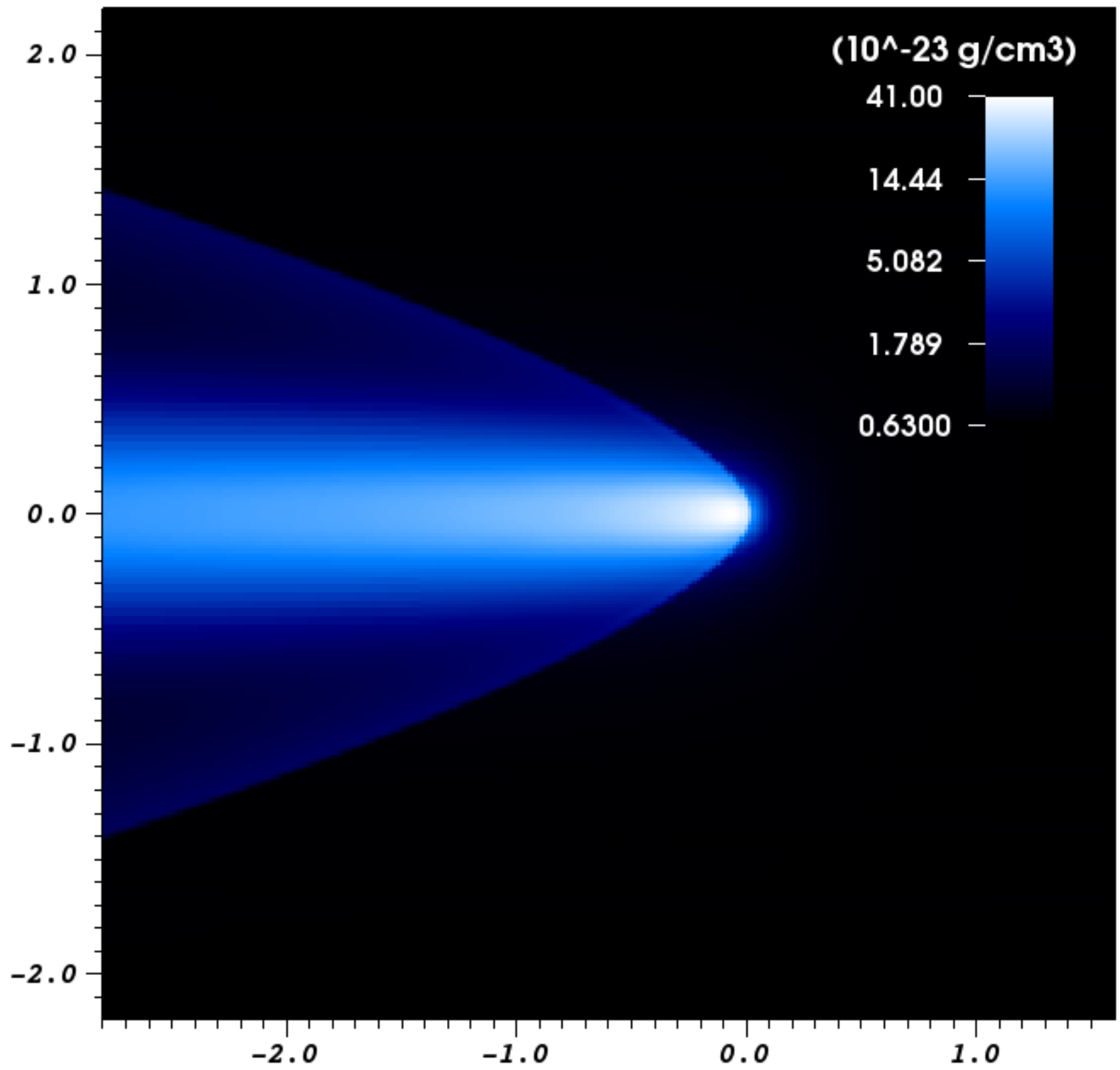} & 
 \includegraphics[width=0.305\linewidth]{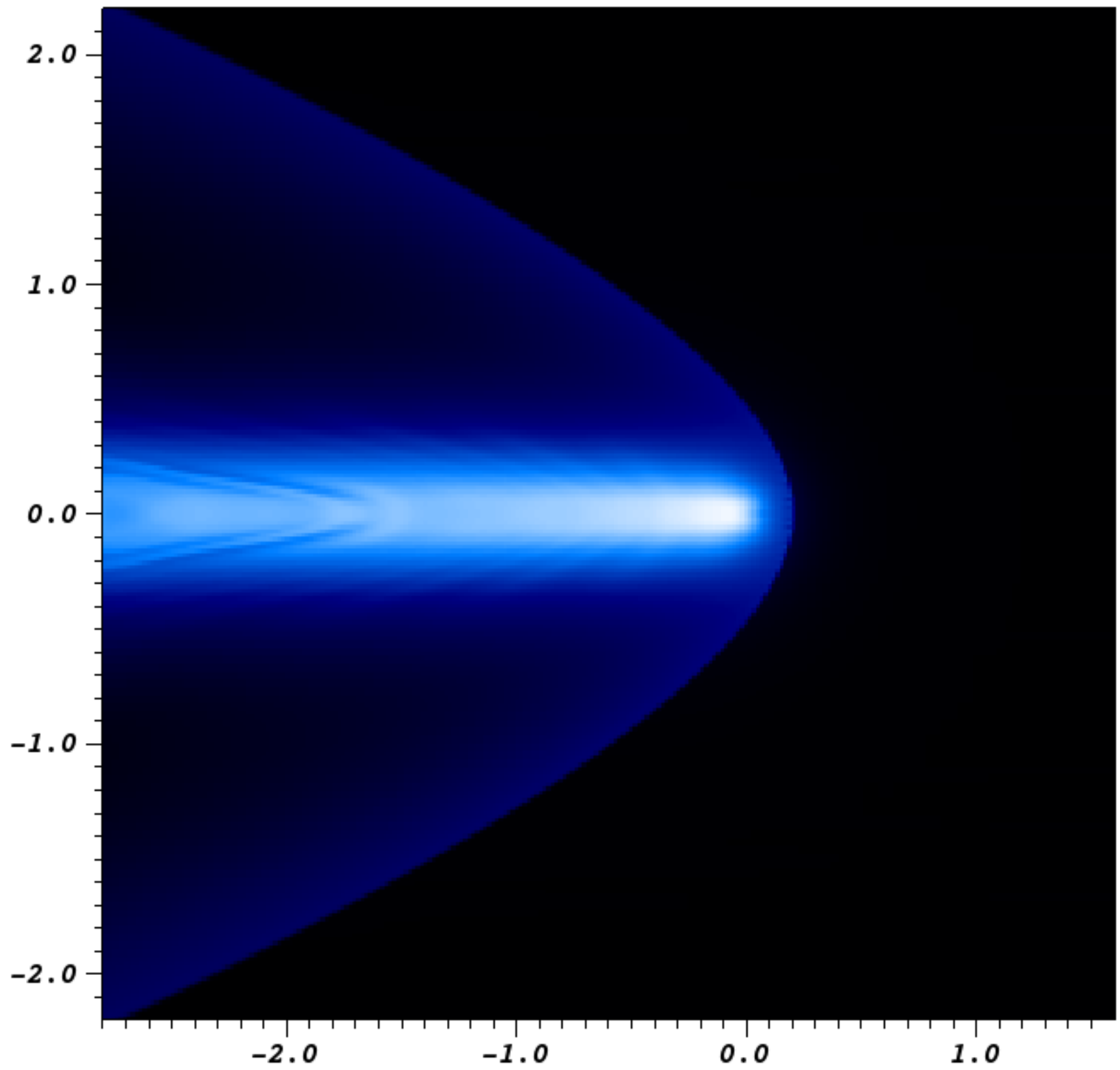} &
 \includegraphics[width=0.305\linewidth]{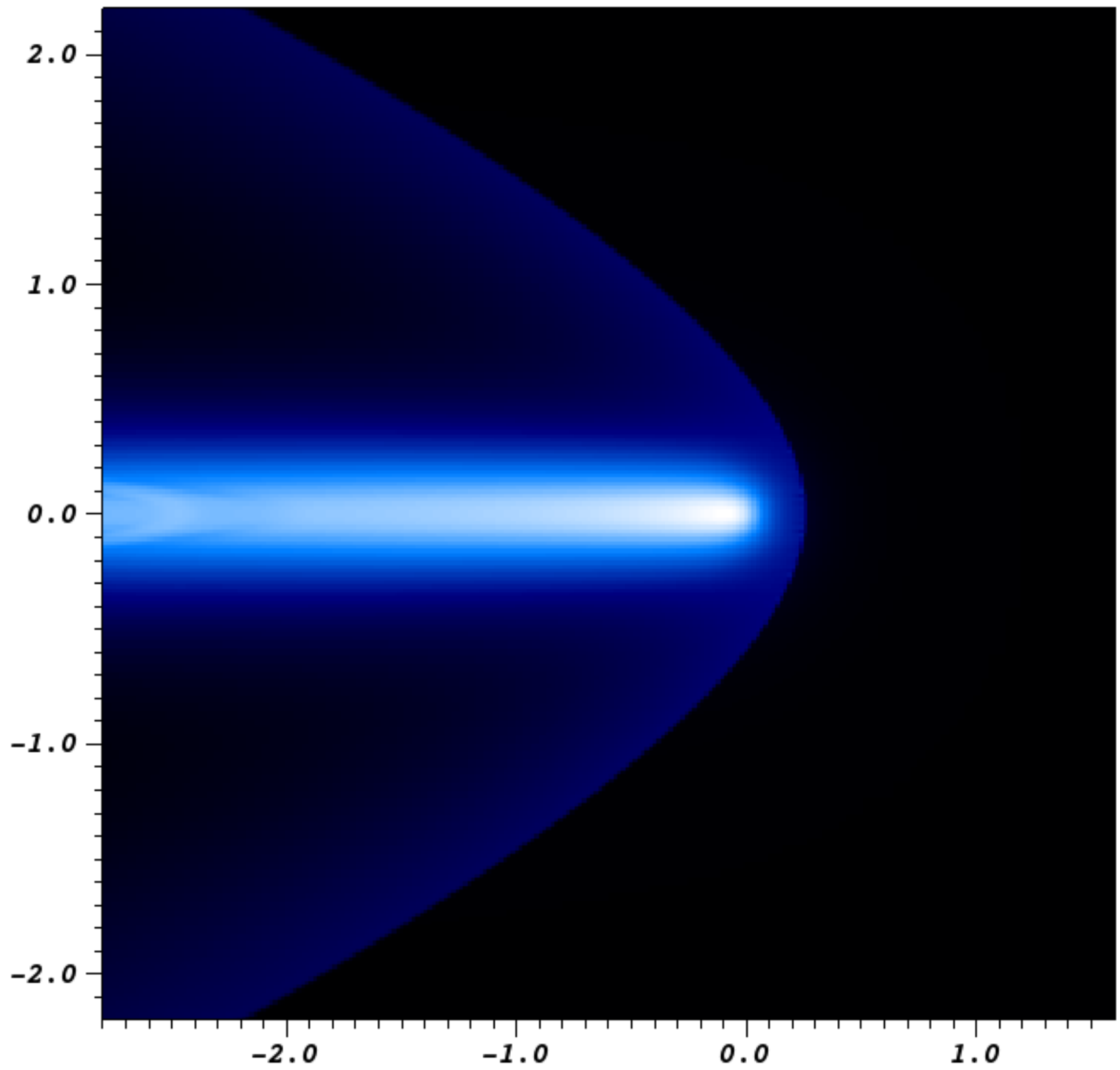} \\
 
 \includegraphics[width=0.305\linewidth]{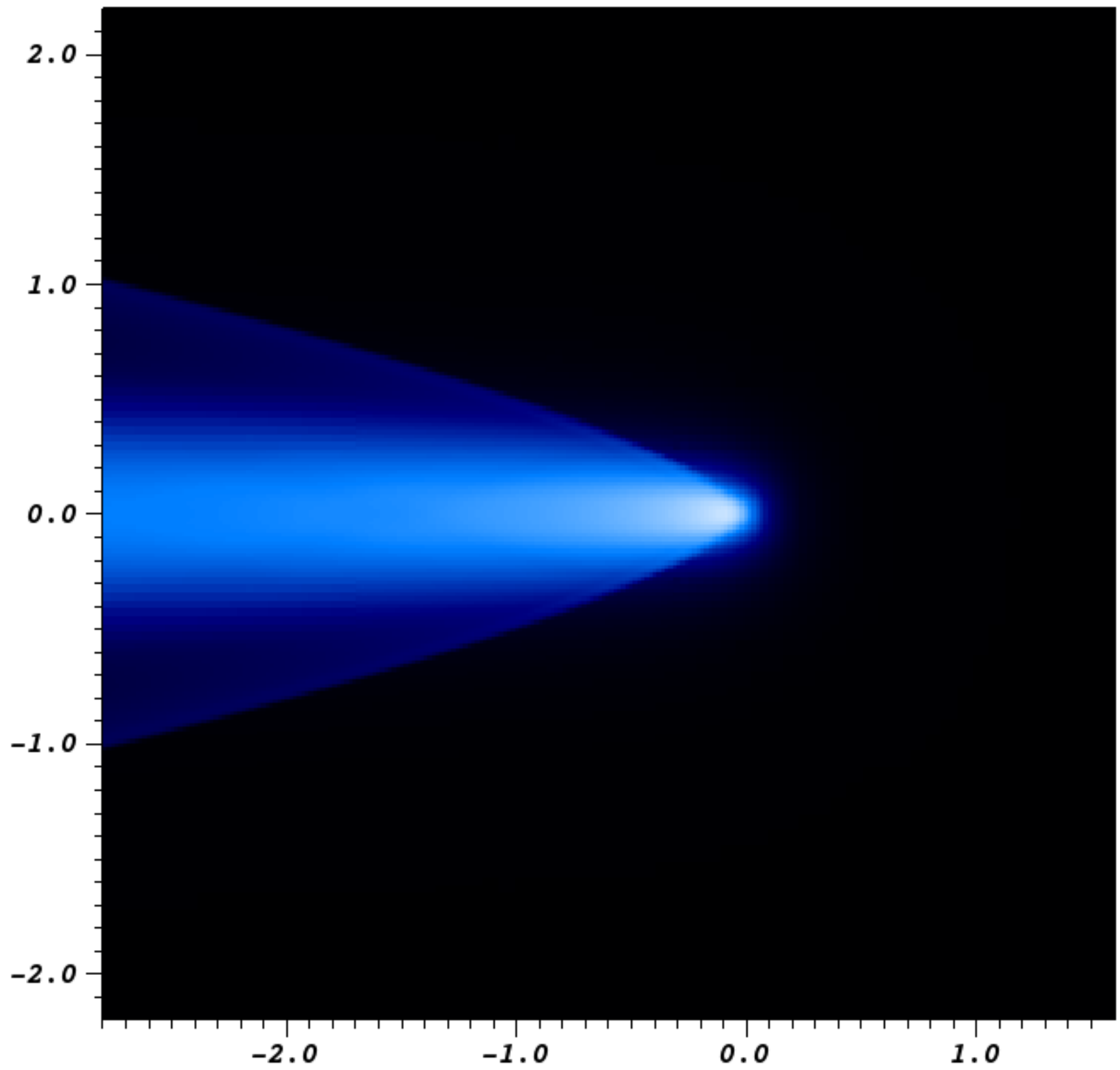} & 
 \includegraphics[width=0.305\linewidth]{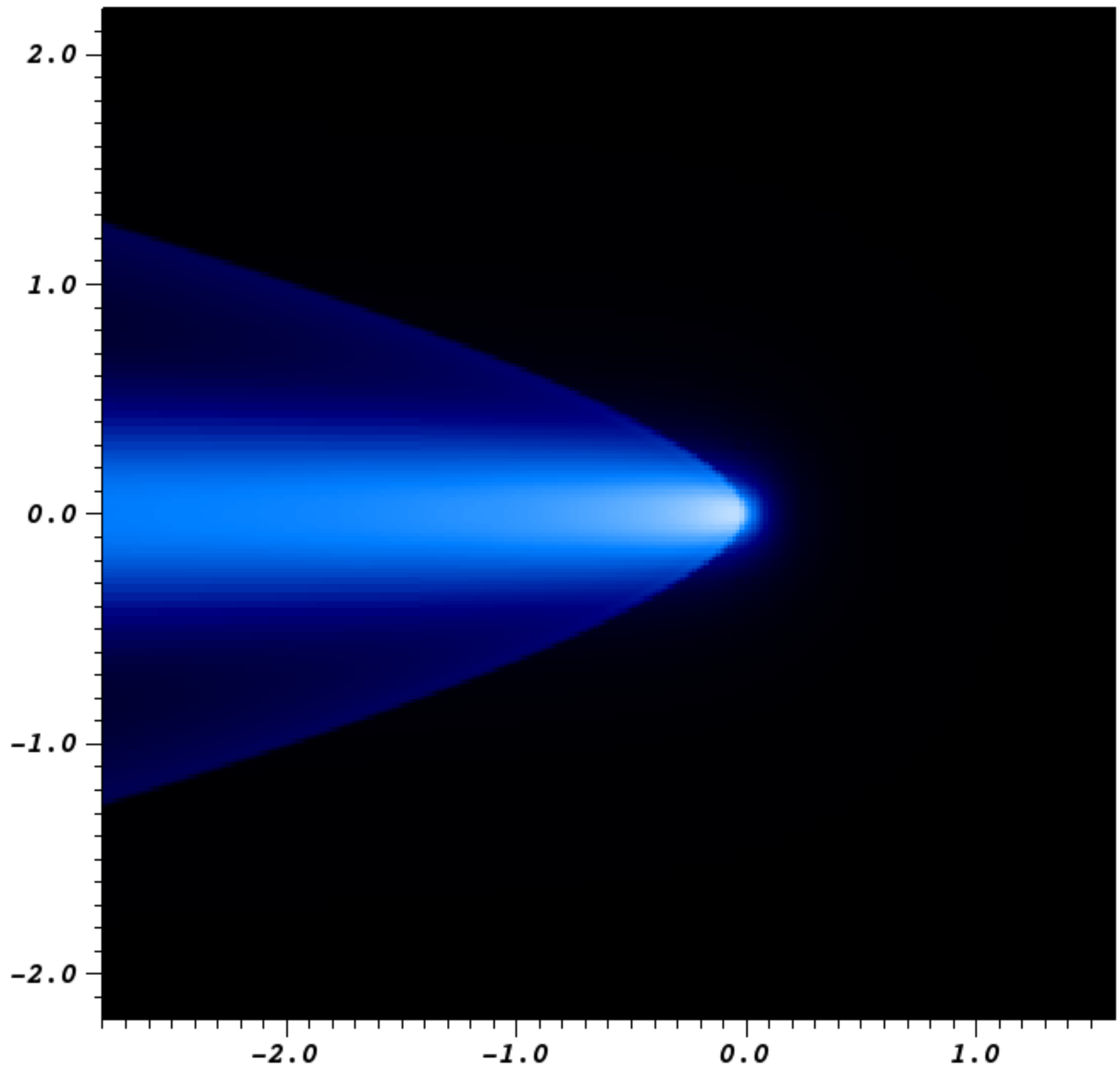} &
 \includegraphics[width=0.305\linewidth]{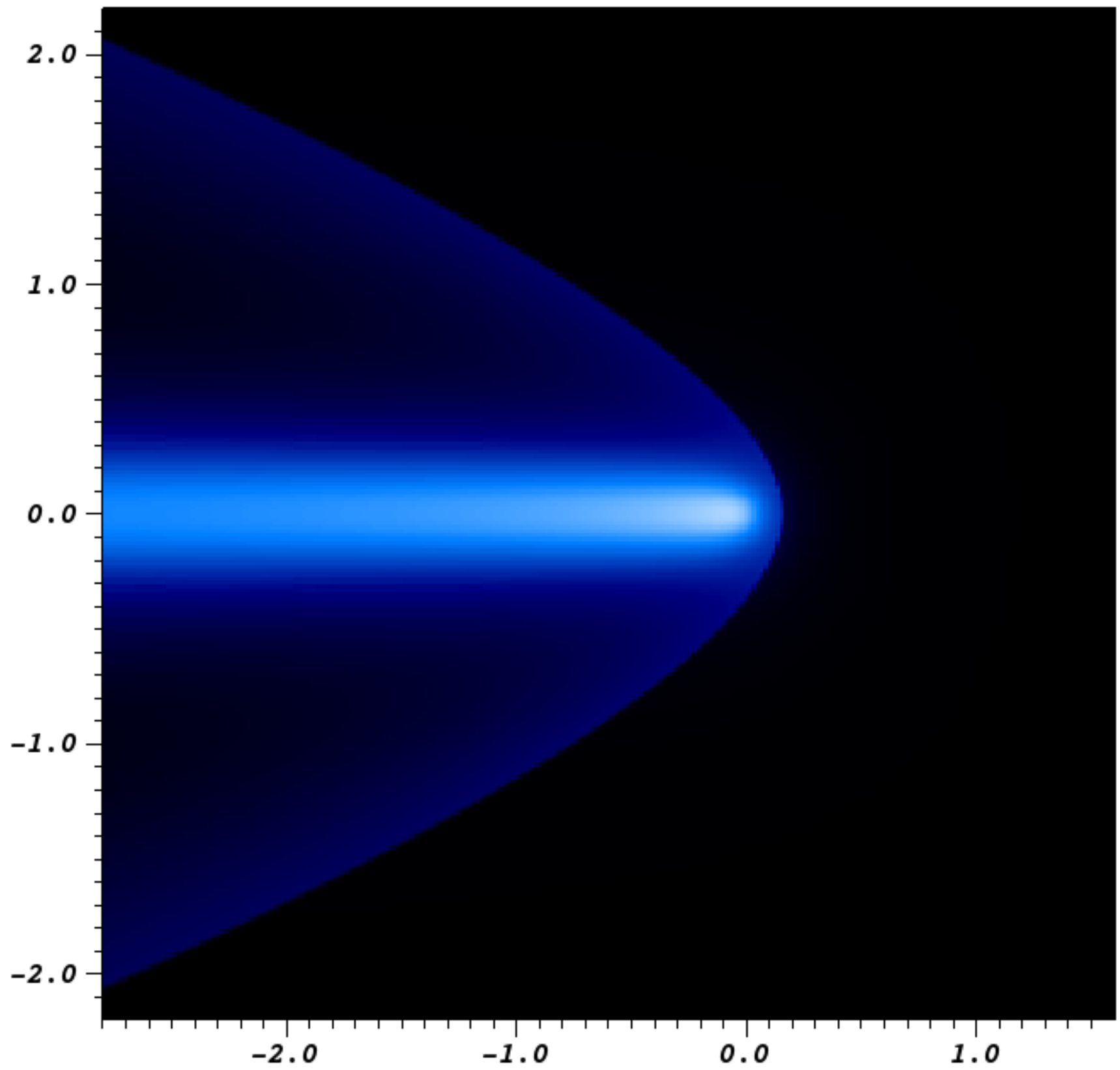} \\
 
 \includegraphics[width=0.305\linewidth]{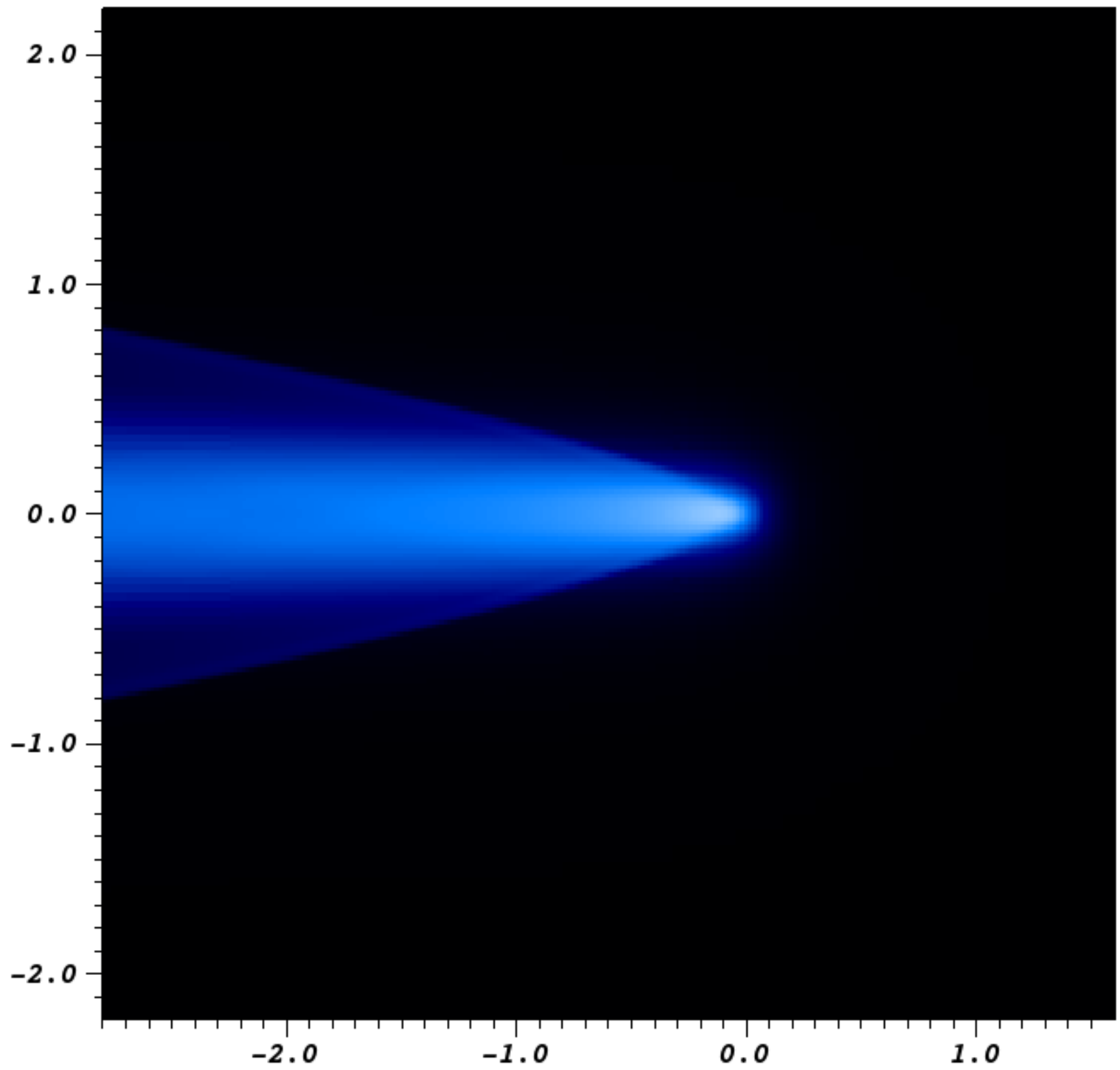} &
 \includegraphics[width=0.305\linewidth]{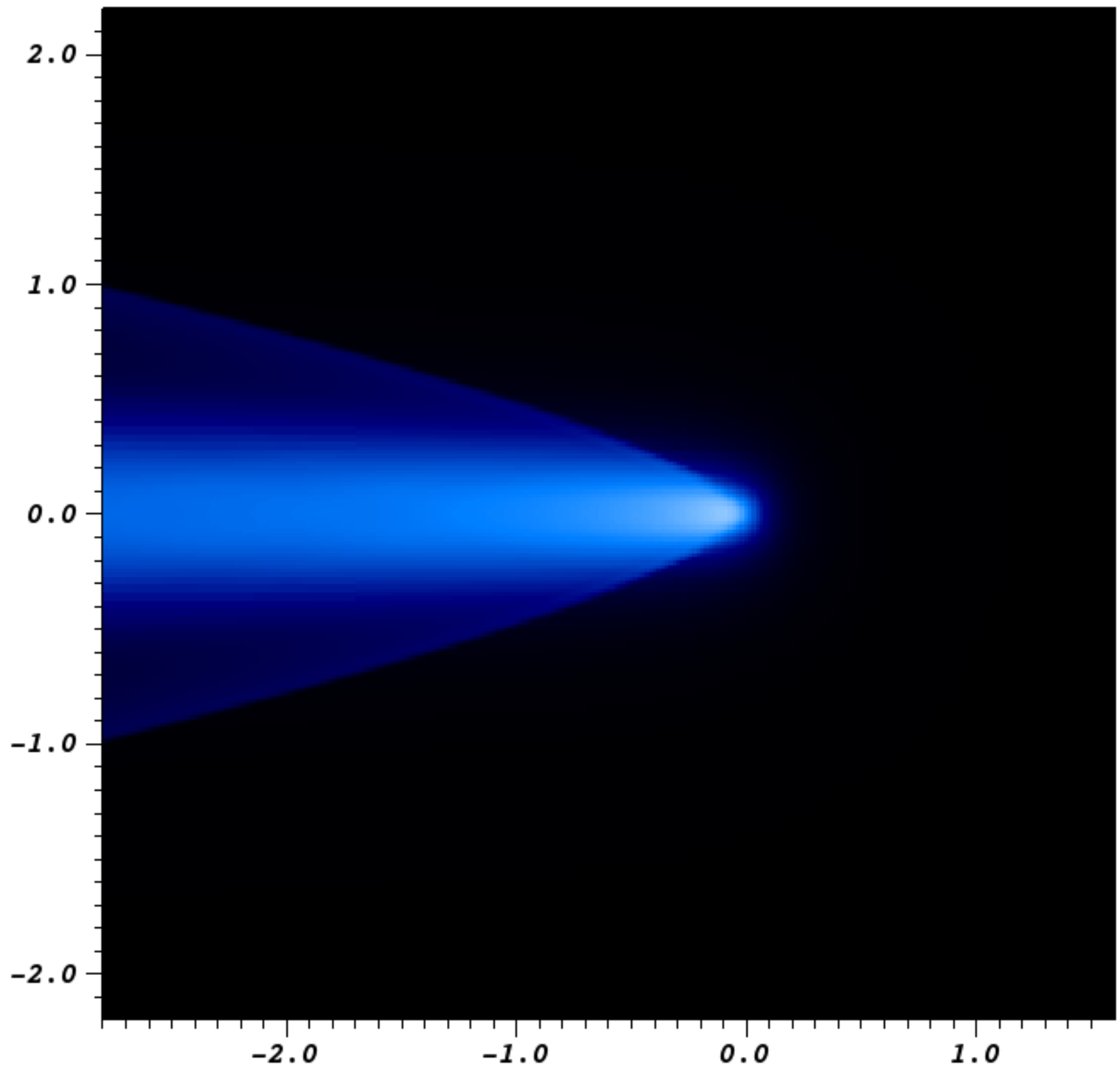} &
 \includegraphics[width=0.305\linewidth]{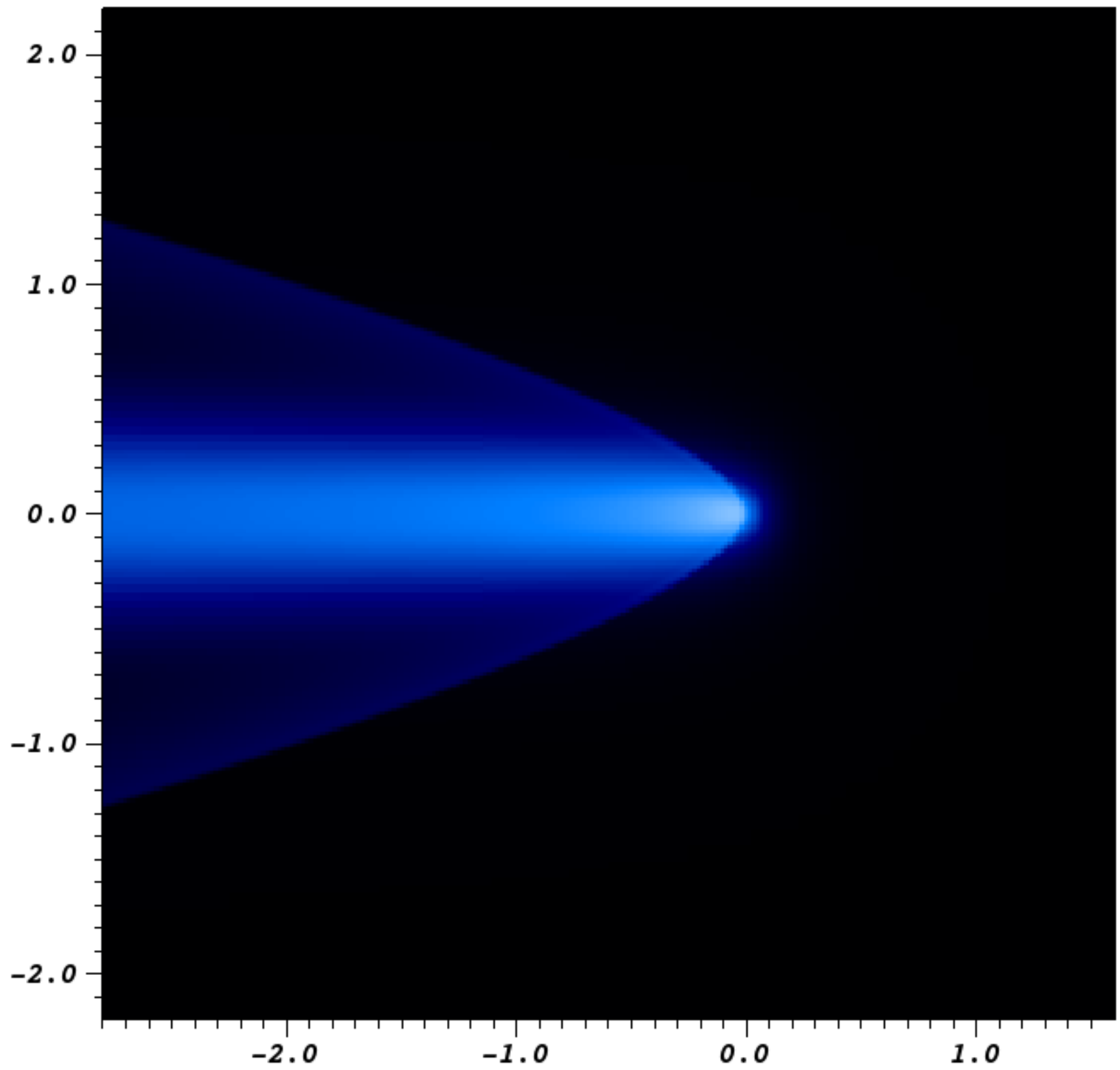} \\
\end{tabular}
\caption{Same as in Fig.~\ref{figsQ1} for $\unit[Q=1.0\times 10^{29}]{s^{-1}}$.}
\label{figsQ2}
\end{figure}

\begin{figure}
\centering
\begin{tabular}{ccc}
 \includegraphics[width=0.305\linewidth]{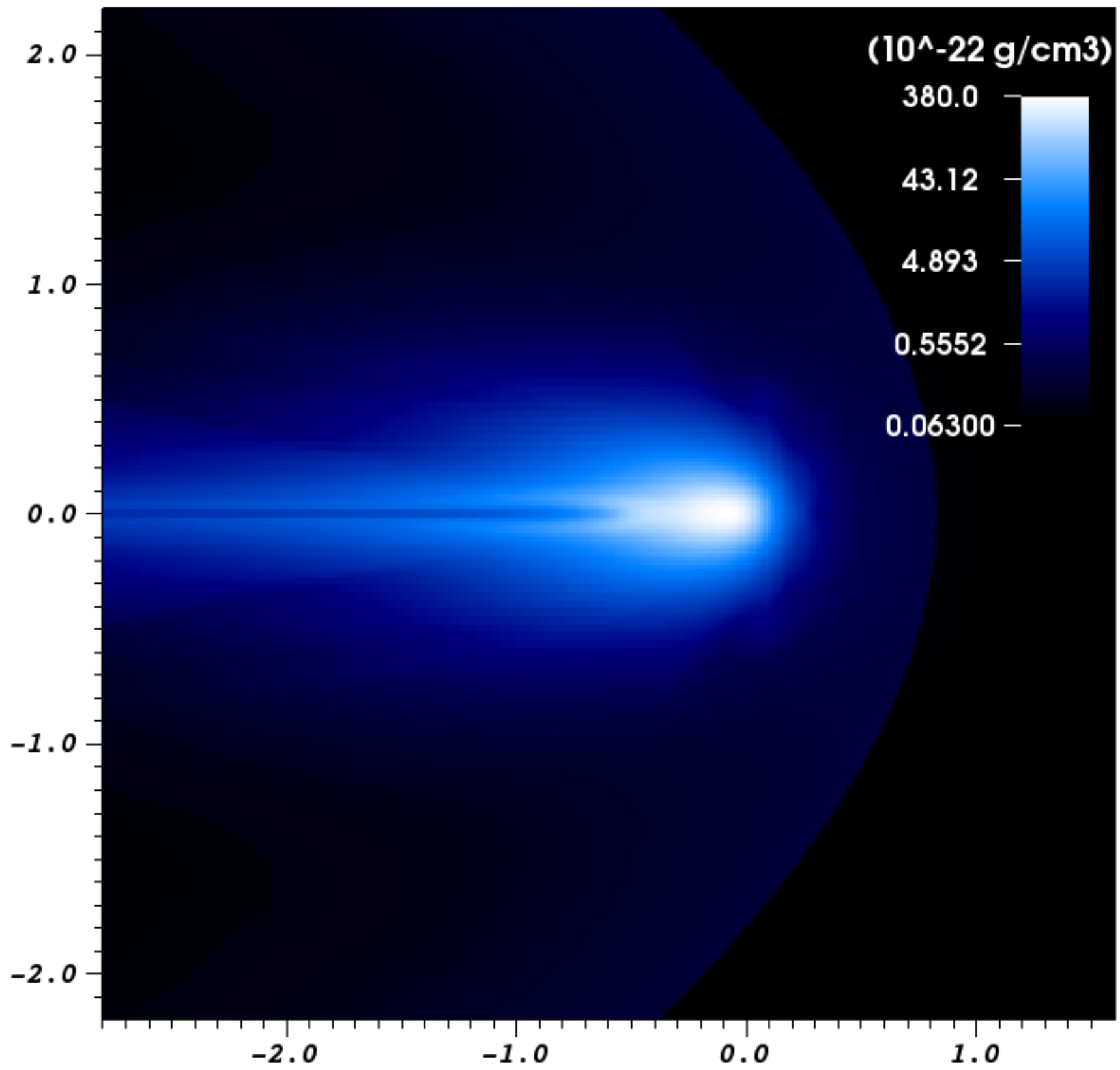} & 
 \includegraphics[width=0.305\linewidth]{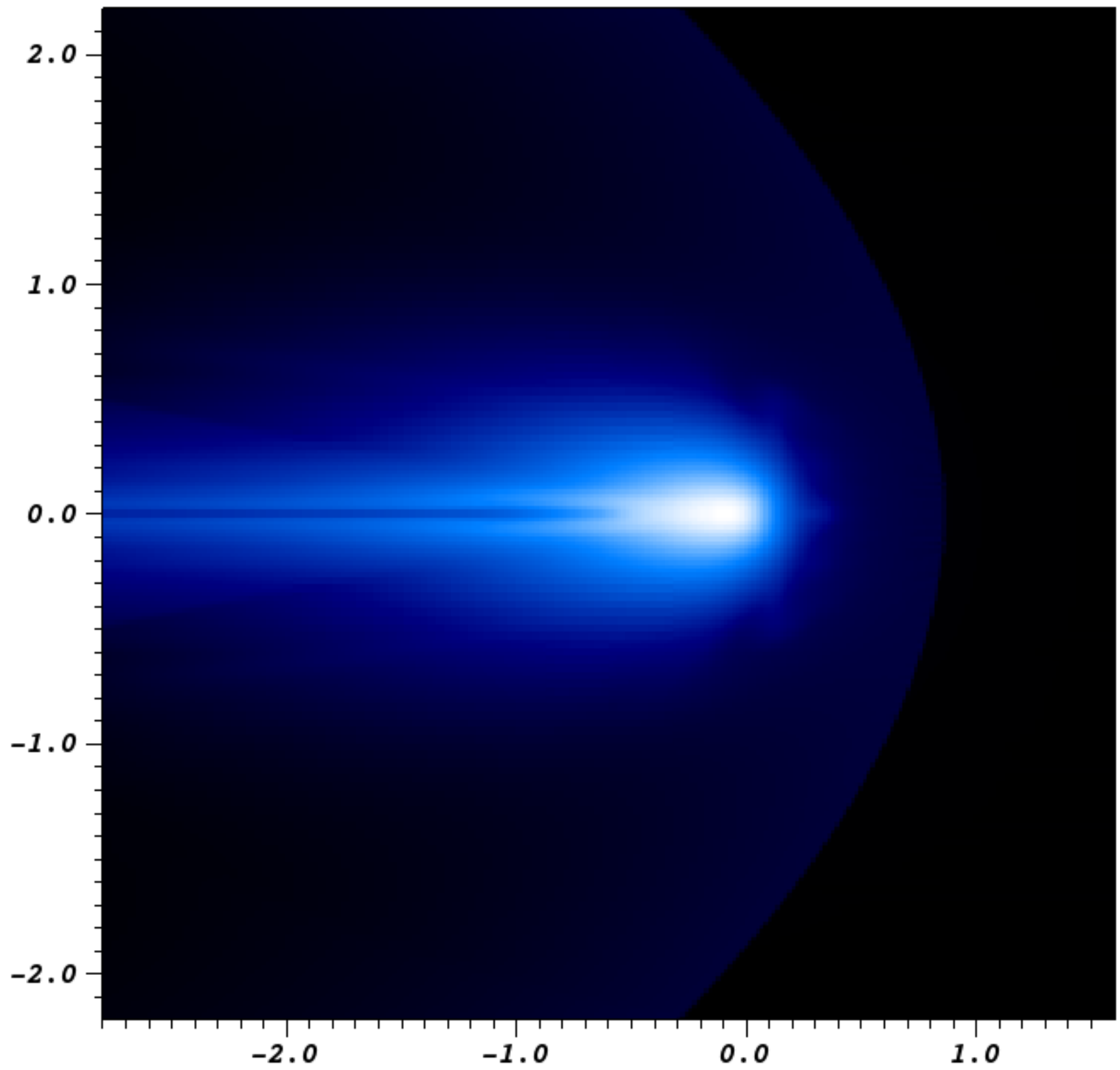} &
 \includegraphics[width=0.305\linewidth]{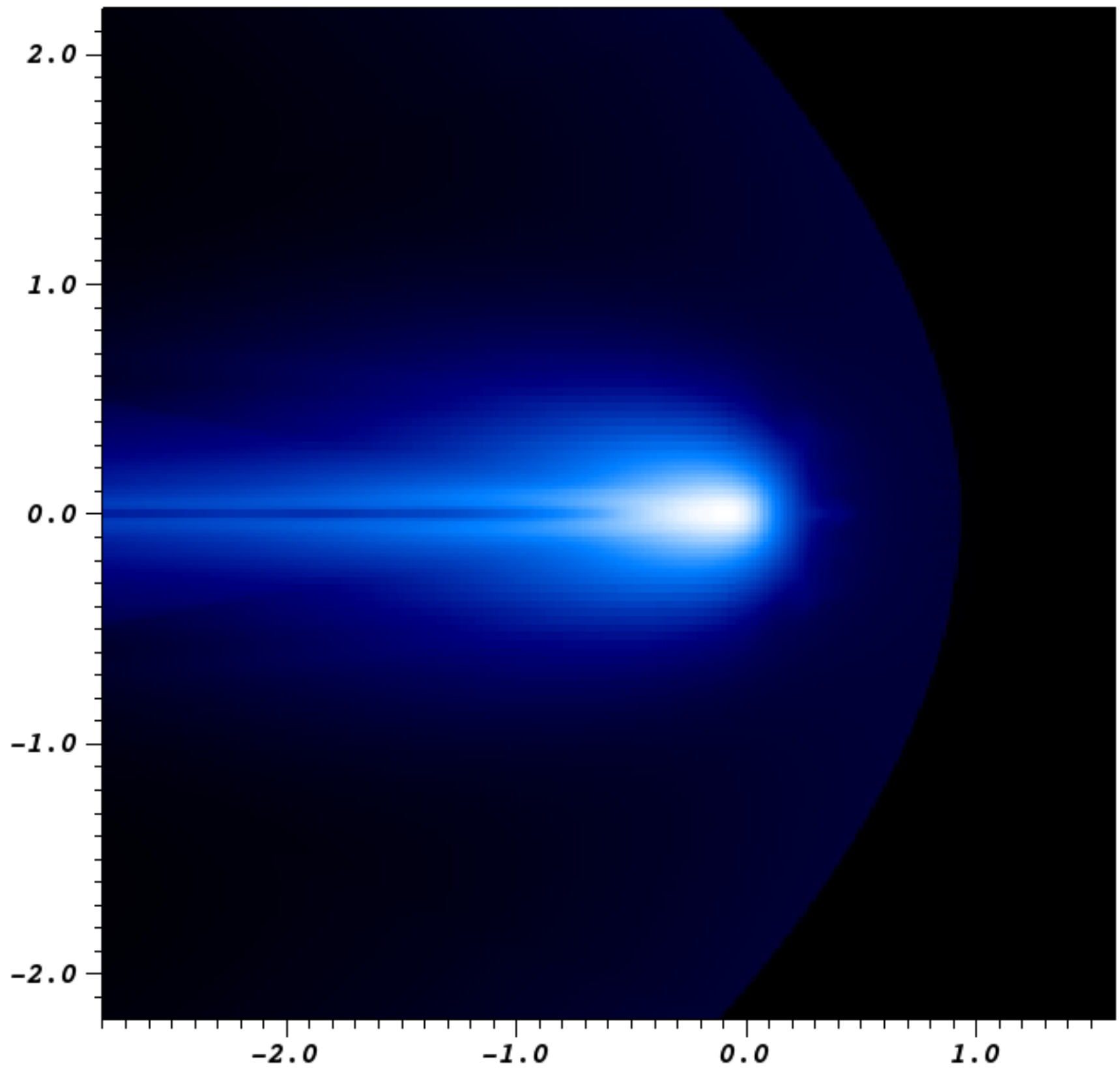} \\
 
 \includegraphics[width=0.305\linewidth]{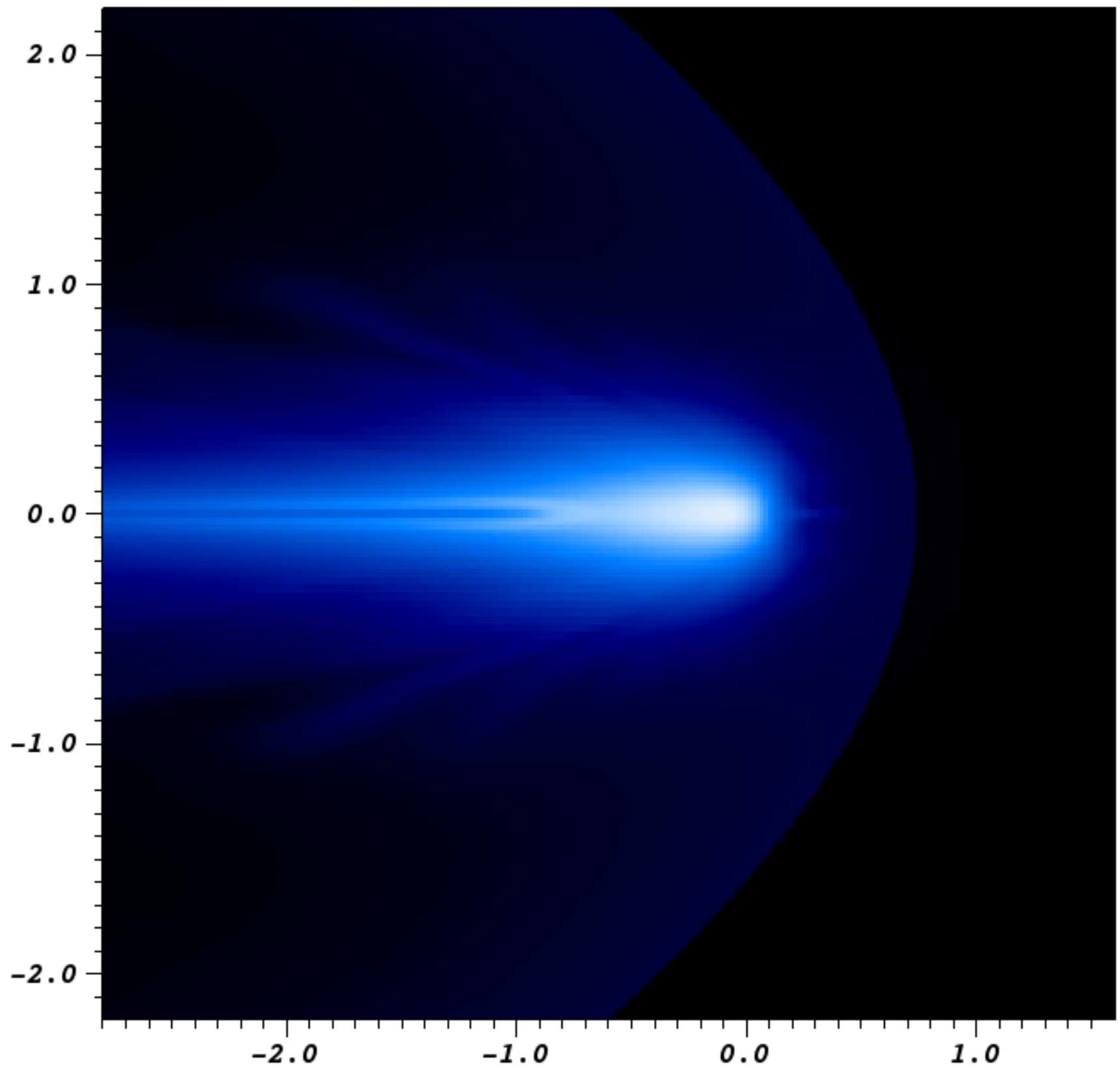} & 
 \includegraphics[width=0.305\linewidth]{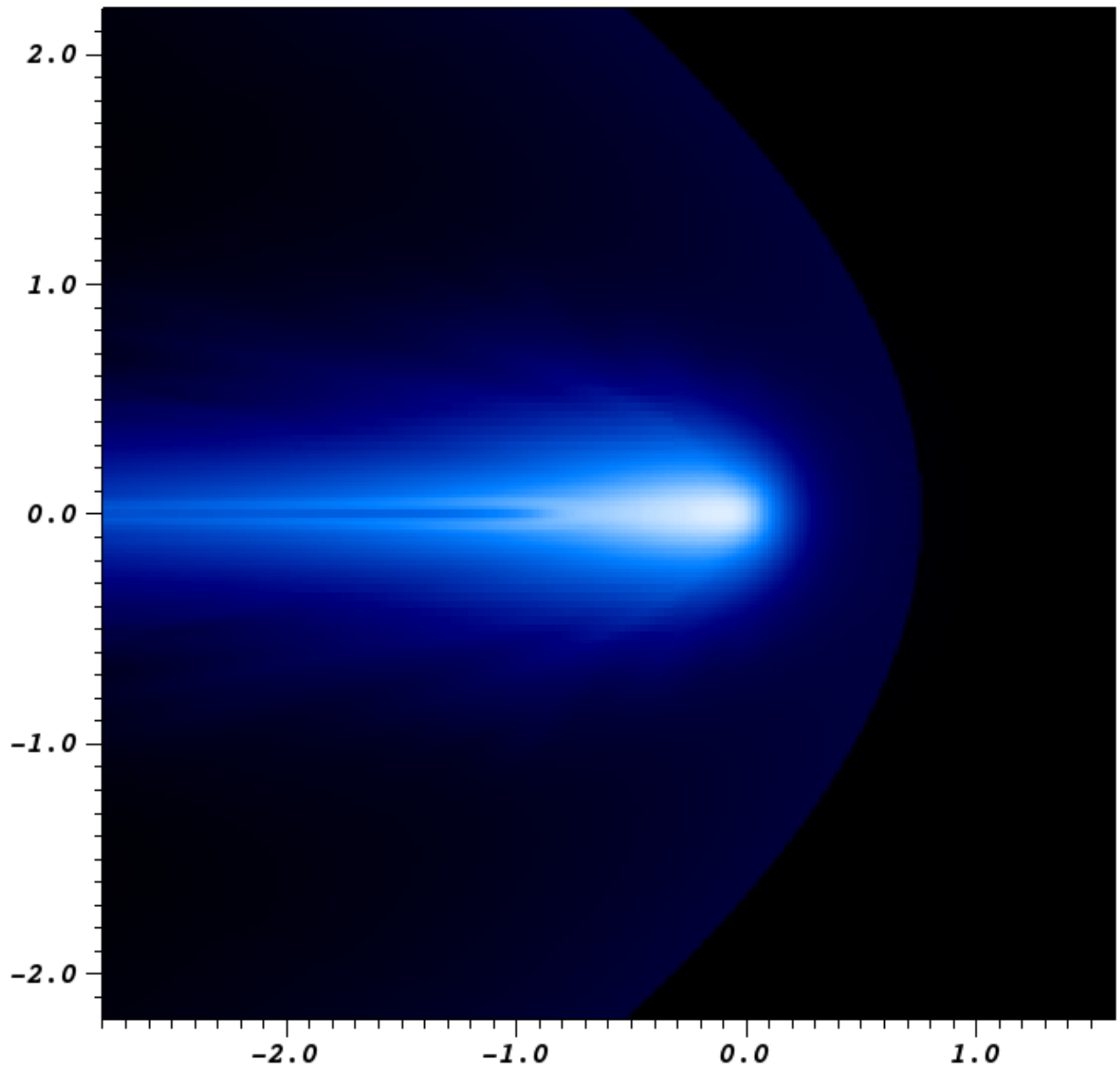} &
 \includegraphics[width=0.305\linewidth]{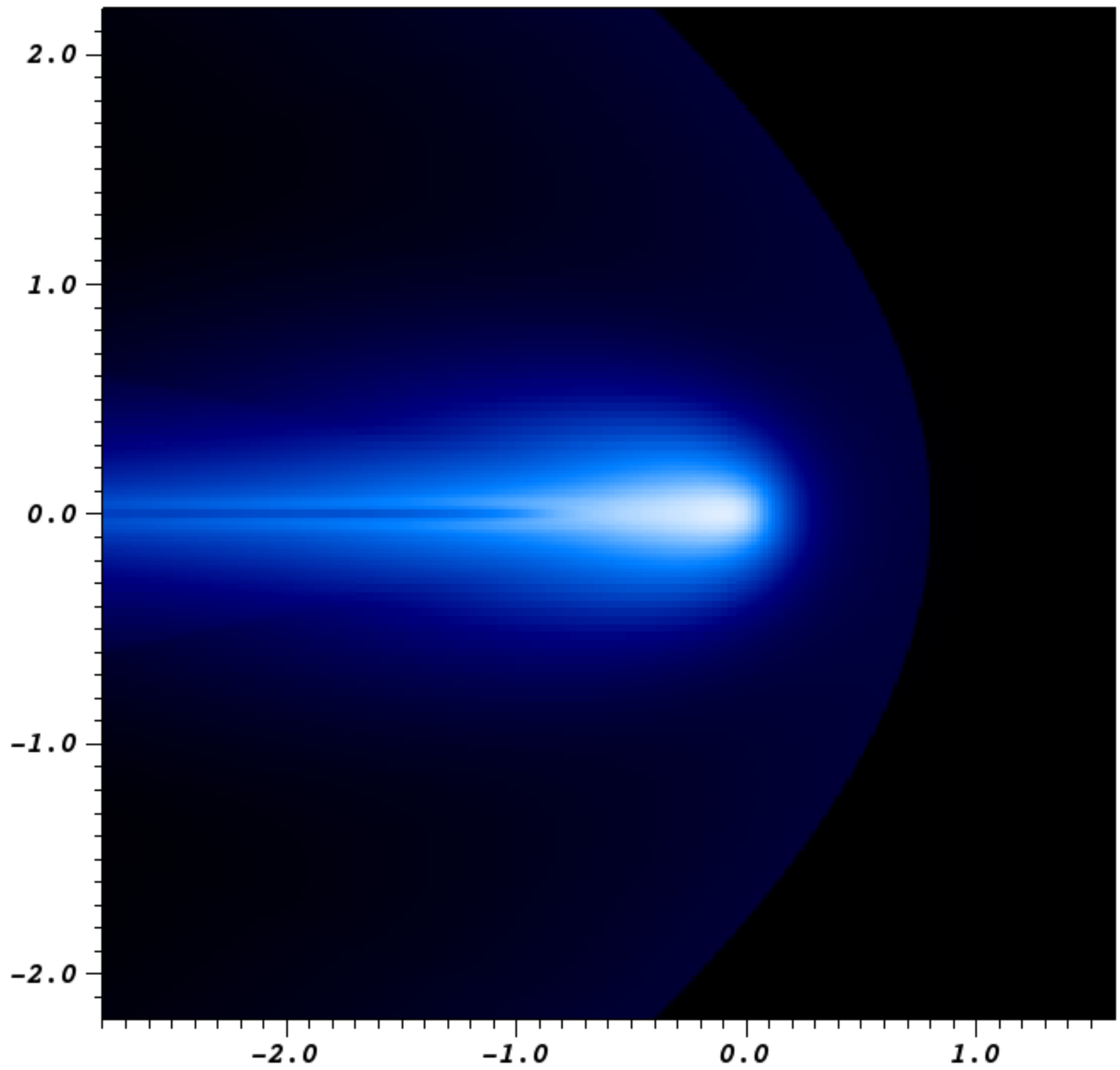} \\
 
 \includegraphics[width=0.305\linewidth]{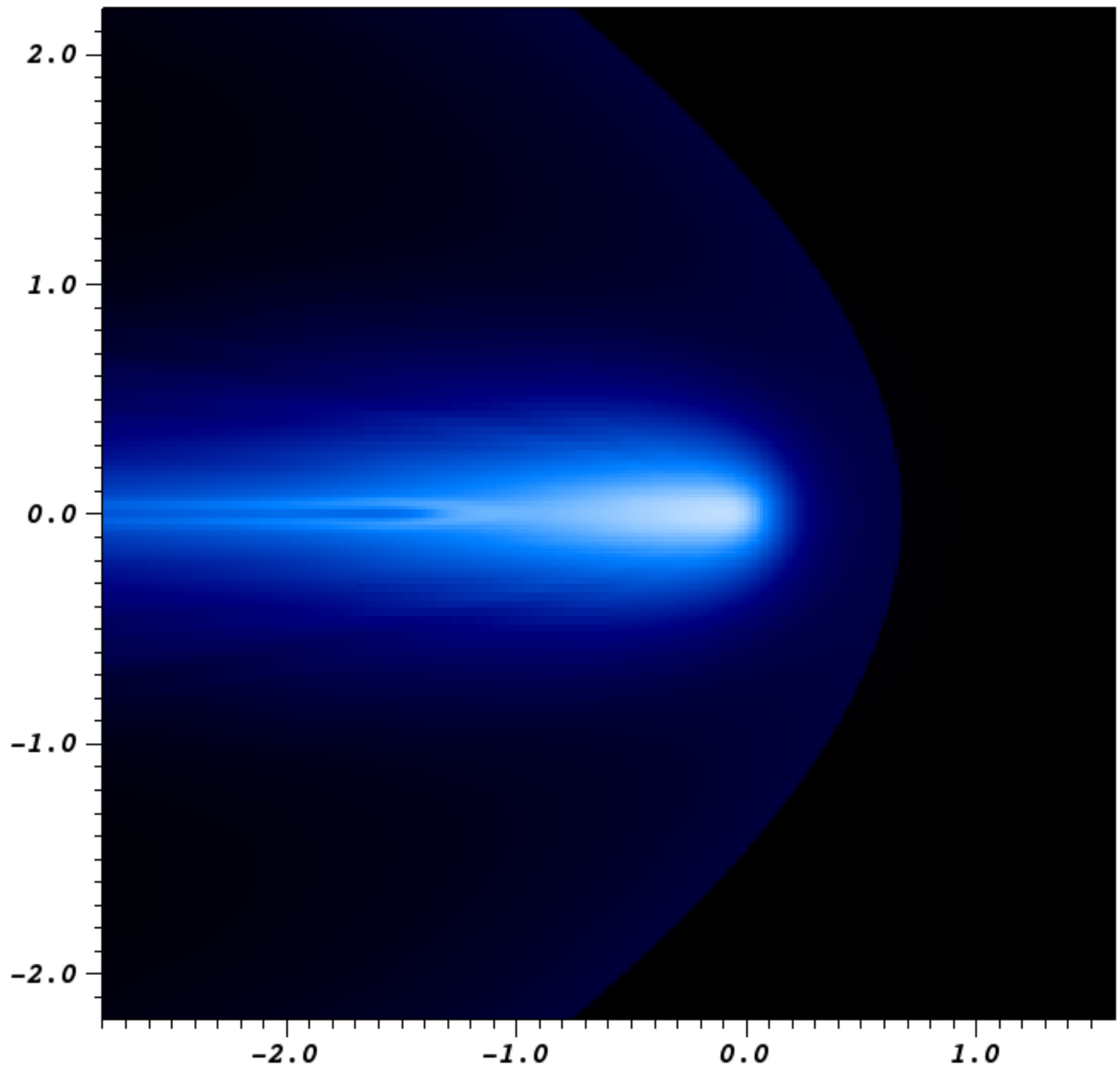} &
 \includegraphics[width=0.305\linewidth]{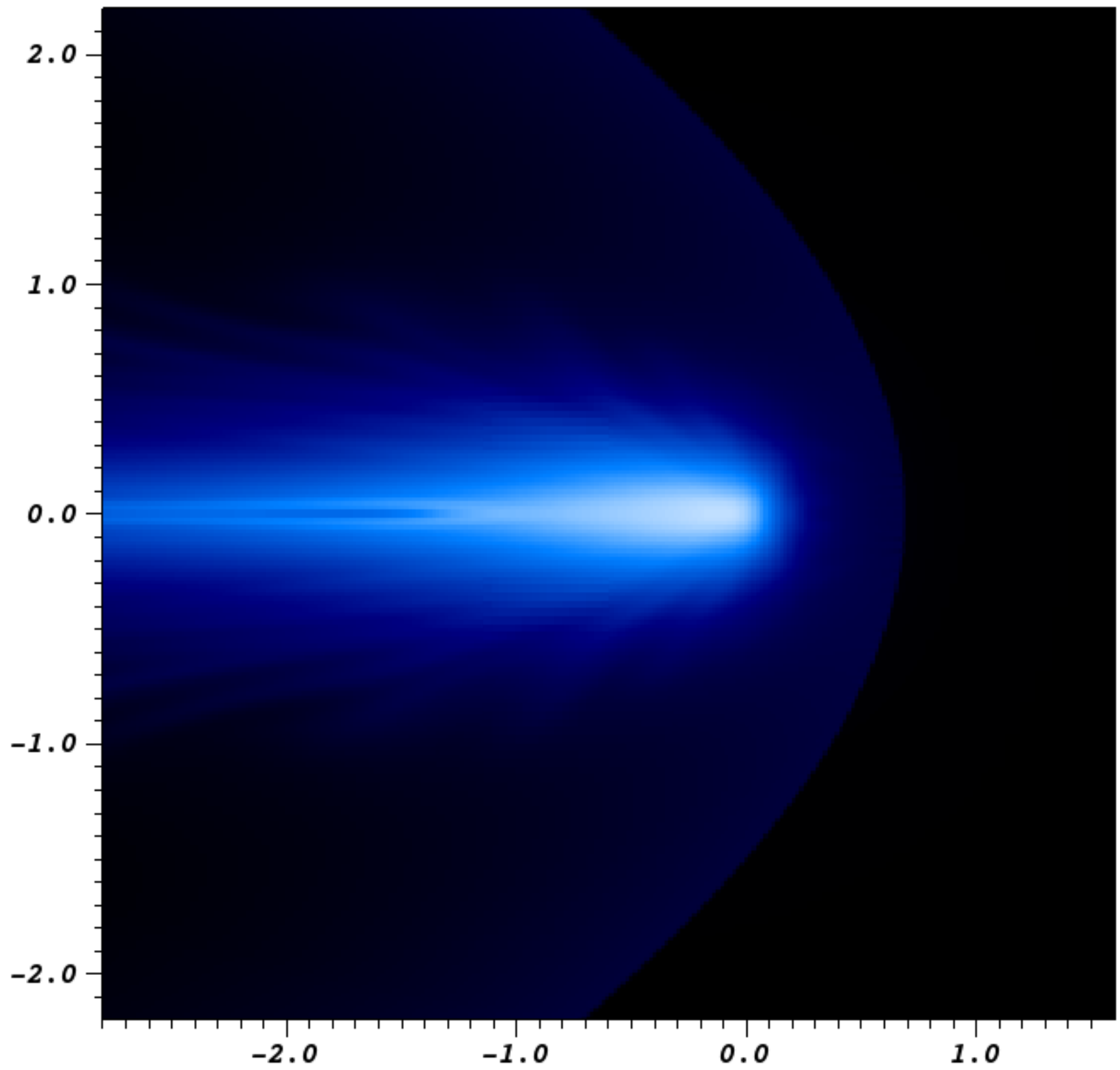} &
 \includegraphics[width=0.305\linewidth]{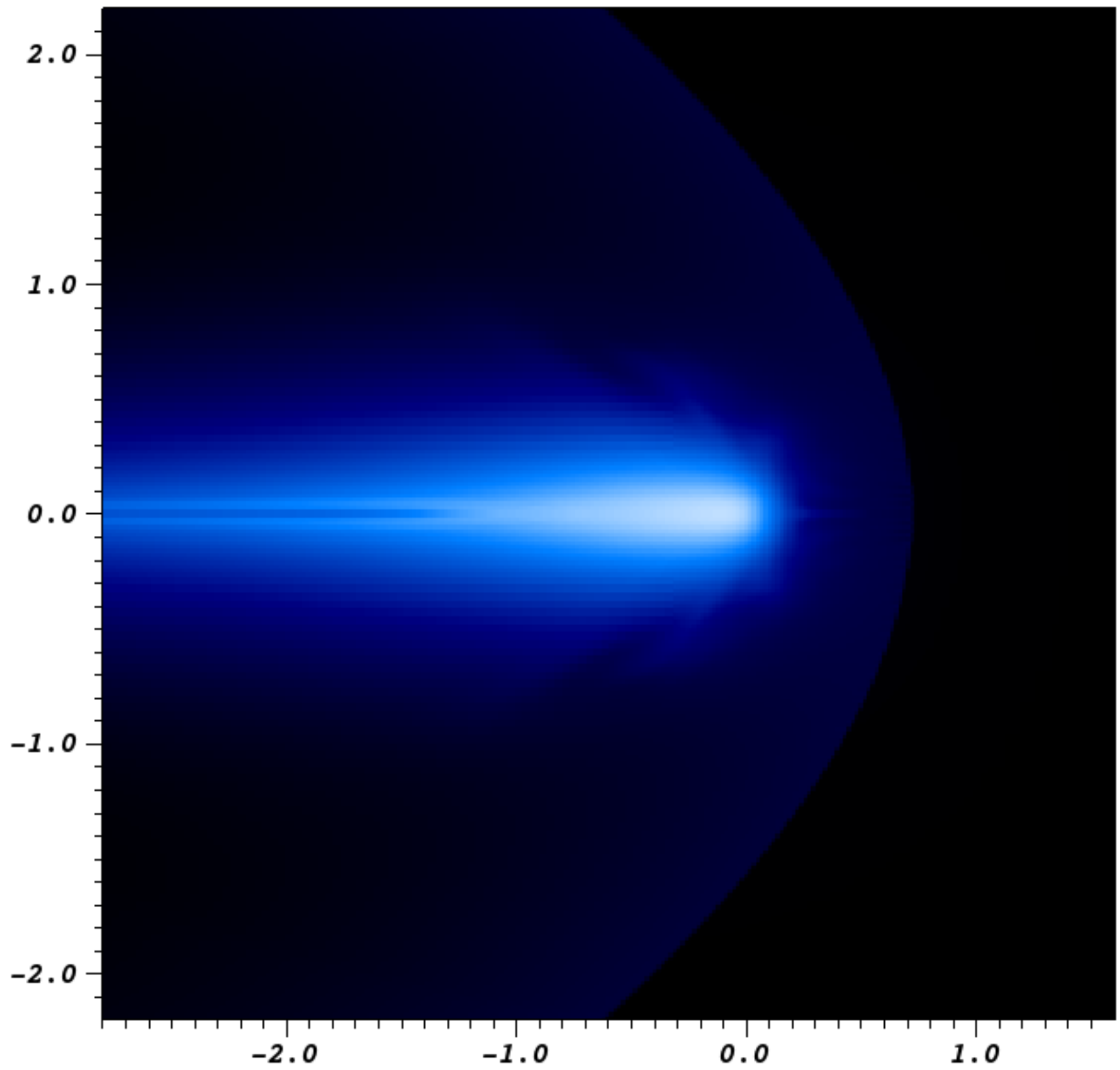} \\
\end{tabular}
\caption{Same as in Fig.~\ref{figsQ1} for $\unit[Q=1.0\times 10^{30}]{s^{-1}}$.}
\label{figsQ3}
\end{figure}

The panels represented in Fig.~\ref{figsQ2} and Fig.~\ref{figsQ3} follow a scheme similar to Fig.~\ref{figsQ1}, but corresponding to $\unit[Q=1.0\times 10^{29}]{s^{-1}}$ and $\unit[Q=1.0\times 10^{30}]{s^{-1}}$.

Observing Fig.~\ref{figsQ1}, one may note the influence of the magnetic field on the formation of the coma and tail of the comet. Indeed, from left to right, the panel show that the magnetic field has the effect of broadening the coma and the bow shock formed from the coma. Besides, note that for higher velocities of the solar wind, the density of the coma and of the tail are smaller. Further, the velocity of the solar wind has the effect of narrowing the bow shock, as may be observed from top to bottom in the three columns of Fig.~\ref{figsQ1}.

The analysis above, concerning the role of the magnetic field and of the velocity of the solar wind, are valid in the case presented in the panel shown in Fig.~\ref{figsQ2}. However, we may note the effect of the increase of $Q$ on the simulation: the tail is thicker and with more complex structure; the densities are higher when compared to the values present in Fig.~\ref{figsQ1} and the bow shock is wider and has higher densities when compared to the case given in Fig.~\ref{figsQ1}.

Particularly, note the structures present in the coma of the profiles shown in the first line of Fig.~\ref{figsQ2}. Such stuctures may be due to the action of the magnetic field and may suggest us a scenario suitable for the onset of MHD instabilities and turbulence.

In Fig.~\ref{figsQ3} we have the highest densities when compared to the previous cases and we may observe the most exhuberant coma and bow shock. Besides, Fig.~\ref{figsQ3} shows interesting characteristics not present in Figs.~\ref{figsQ1}-\ref{figsQ2} such as the filament-like structures around the coma. As the parameters of the solar wind are the same as in the previous cases, we deduce that the formation of such structures is related to the rate $Q$. More specifically, in the scenarios considered here, rates $Q\gtrsim \unit[10^{30}]{s^{-1}}$ are suitable for the formation of the filaments. As in the case of the structures in Fig.~\ref{figsQ2}, such filaments may indicate us the onset of instabilities.

Further, in Fig.~\ref{figsQ3} we may observe the influence of the velocity and magnetic field of the solar wind on the characteristics of the comet, though such an influence is not as strong as in the previous cases shown here. Such a fact suggests us that for higher values of $Q$, this rate becomes more important on the definition of the characteristics of the coma and tail.

\section{Conclusions}

\noindent
We simulated the interaction between a comet and the solar wind. The comet was treated as a spherically symmetric source of ions.

We considered several scenarios, with different values for the velocity $v$ and the magnetic field $B_z$ of the solar wind, as well as different values for the gas production rate $Q$ of the comet. Particularly, we used three different values for $Q$, $v$ and $B_z$, yielding 27 different scenarios. In the simulations, we observed that the magnetic field has the effect of widening the coma and the bow shock; while for higher $v$ we have smaller densities of the coma.

On the other hand, the rate $Q$ plays an important role in the sense that, for higher $Q$, we have more exhuberant comas, tails and bow shocks. Besides such characteristics, a high value of $Q$ produced the filaments around the coma observed in Fig.~\ref{figsQ3}.

We observed that the physical scenarios simulated here are suitable for the onset of MHD phenomena such as turbulence and instabilities. In fact, the presence of the structures observed in the tail of the comet in Fig.~\ref{figsQ2} and the filaments observed in Fig.~\ref{figsQ3} are suggestive of such phenomena.

In forthcoming papers we will apply the computational scheme shown here in the investigation of particular phenomena occurring in the coma and tail, such as the filaments commented above. The aim will be the study of MHD instabilities and turbulence, which will provide us a better knowledge on the mechanisms involved in the interaction of the solar wind with the cometary atmospheres. Besides, we will consider the case where the solar wind have particular characteristics, such as the presence of shocks and Alfv\`{e}n waves.

\section*{Acknowledgments}

EFD Evangelista acknowledges the Brazilian agency CNPq, grant 300089/2016-3 (PCI INPE).
O Mendes, MO Domingues and OD Miranda acknowledge MCTI/FINEP/INFRINPE-1 (grant 01.12.0527.00), the Brazilian agencies CNPq (grants 306038/2015-3; 312246/2013-7), FAPESP (grant 2015/25624-2), CAPES for financial support. Solar wind data were provided by GSFC/SPDF with the OMNIWeb interface. FLASH was in part developed by the DOE NNSA-ASC OASCR Flash Center at the University of Chicago.

\end{document}